\documentclass[twocolumn,pra,showpacs,superscriptaddress]{revtex4-1}

\usepackage[T1]{fontenc}
\usepackage{color}
\usepackage{amsmath}
\usepackage{amssymb}
\usepackage{graphicx}
\usepackage{esint}
\usepackage[unicode=true,pdfusetitle,
 bookmarks=true,bookmarksnumbered=false,bookmarksopen=false,
 breaklinks=false,pdfborder={0 0 1},backref=false,colorlinks=true]
 {hyperref}
\hypersetup{
 citecolor=blue, urlcolor=blue}



\begin{document}

\title{Non-Markovianity in a non-thermal bath}

\date{\today}

\author{Sheng-Wen Li}

\author{Moochan B. Kim}

\affiliation{Texas A\&M University, College Station, TX 77843}

\author{Marlan O. Scully}

\affiliation{Texas A\&M University, College Station, TX 77843}

\affiliation{Princeton University, Princeton, NJ 08544}

\affiliation{Baylor University, Waco, TX 76798}
\begin{abstract}
We study the dynamics of an open quantum system interacting with a
non-thermal bath. Here, ``non-thermal'' means that the bath modes
do not need to have the same temperature, but they have an effective
temperature distribution. We find that, when a quantum system is interacting
with such a non-thermal bath far from thermal equilibrium, it is no
longer proper to use any coarse-grained Markovian description for
the system, even when their coupling strength is quite weak. Especially,
when there is coherent transition with strong interference strength
in the quantum system, the Markovian master equation would bring in
a serious problem of negative probability. After we consider some
proper non-Markovian corrections, the problem can be naturally resolved. 
\end{abstract}

\pacs{03.65.Yz, 05.30.-d, 05.70.Ln}

\maketitle

\section{Introduction}

The property of the bath has an critical influence to the dynamical
behavior of an open quantum system. For example, common bath could
give rise to decoherence-free subspace and dark state, while independent
baths could not \cite{duan_preserving_1997,zanardi_noiseless_1997,lidar_decoherence-free_1998,scully_quantum_1997};
Squeezed baths could afford a thermal machine beyond the Carnot efficiency,
which is the upper limit for canonical thermal baths \cite{rosnagel_nanoscale_2014}.
These physical effects all result from the special properties of their
particular baths.

When we study the dynamics of an open quantum system, we usually consider
it to be interacting with a bath in thermal equilibrium, which is
described by the canonical Gibbs state, 
\begin{equation}
\rho_{B}=\frac{1}{{\cal Z}_{B}}\exp[-\frac{1}{T}\,\hat{H}_{B}].\label{eq:therm-state}
\end{equation}
 Here $\hat{H}_{B}$ is the Hamiltonian of the bath and $T$ is the
temperature (we set the Boltzmann constant $k_{B}=1$). Thermal equilibrium
is an idealistic physics model, and non-thermal baths also exists
quite widely in realistic physical world. The different bath modes
in a non-thermal bath do not have to share the same temperature $T$
as Eq.\,(\ref{eq:therm-state}), but they could have an effective
temperature distribution \cite{alicki_non-equilibrium_2015}\@. In
the studies of biology systems whose environments are usually quite
complicated \cite{dorfman_photosynthetic_2013,creatore_efficient_2013,olsina_can_2014},
this non-thermal bath model could provide a description more close
to the realistic situation.

In this paper, we study the dynamics of an open quantum system in
such a non-thermal bath. We find that, if the bath state is far from
thermal equilibrium, the relaxation of the bath will be quite slow.
And it is no longer appropriated to use constant decay rates to describe
the evolution of the open system. Due to the slow relaxation of the
bath, the decay rates of the system also vary slowly, and sometimes
they could even become negative. As the result, the dynamics of the
open system shows typical non-Markovian features, and we cannot describe
the system by a coarse-grained Markovian master equation as before,
although the coupling strength between the system and the bath is
very weak. This is different from most of previous studies where the
non-Markovianity is usually caused by the strong system-bath coupling
\cite{hu_quantum_1992,breuer_theory_2002,ma_entanglement_2012,yang_master_2013,li_approach_2014,weiss_quantum_2012}. 

Especially, we are interest in the case when the quantum system has
coherent transitions, also well-known as the Fano-Agarwal interference
\cite{agarwal_quantum_1974,zhu_quantum-mechanical_1995,li_quantum_2010}.
In this case, when the coherent transition has maximum interference,
the Markovian master equation would bring in a serious negative probability
problem, even when the system-bath coupling strength is quite weak.
Indeed, this negative probability problem resulted from the coherent
transition also exists when the bath is a thermal one, but the negative
value is negligibly small and it only lasts for very short time. However,
when the bath is very far from thermal equilibrium, this negative
probability is significantly large and exists for very long time,
which is intolerable. We find that if the non-Markovian correction
is taken into consideration, this negative probability problem could
be naturally resolved.

This paper is organized as follows. In Sec.\,II, we give a brief
discussion about the property of a non-thermal bath. Then we study
the time-dependent evolution of a three-level system interacting with
a non-thermal bath in Sec.\,III, and discuss the validity of the
approximations made for the master equation. We will show that there
will be negative probability probability when the coherent transition
has maximum interference. In Sec.\,IV, we show that this negative
probability problem can be resolve by introducing non-Markovian correction.
We finally draw conclusion in Sec.\,V.

\section{Non-thermal bath}

In this section, we discuss the properties of the non-thermal bath
we are going to study in this paper. Usually when we study the dynamics
of an open quantum system, we consider that it is coupled with a boson
bath in thermal equilibrium, which is described by the thermal state
\begin{equation}
\rho_{B}=\frac{1}{{\cal Z}_{B}}e^{-\hat{H}_{B}/T}=\frac{1}{{\cal Z}_{B}}\exp[-\frac{1}{T}\sum_{k}\omega_{k}\hat{b}_{k}^{\dagger}\hat{b}_{k}],
\end{equation}
where $\hat{H}_{B}=\sum\omega_{k}\hat{b}_{k}^{\dagger}\hat{b}_{k}$
is the self Hamiltonian of the boson bath, and $T$ is the temperature.
When the temperature $T\rightarrow0$, the bath tends to the vacuum
state, $\rho_{B}\rightarrow|\mathsf{vac}\rangle\langle\mathsf{vac}|$,
and there is no excitation in each boson mode.

\begin{figure}
\includegraphics[width=0.95\columnwidth]{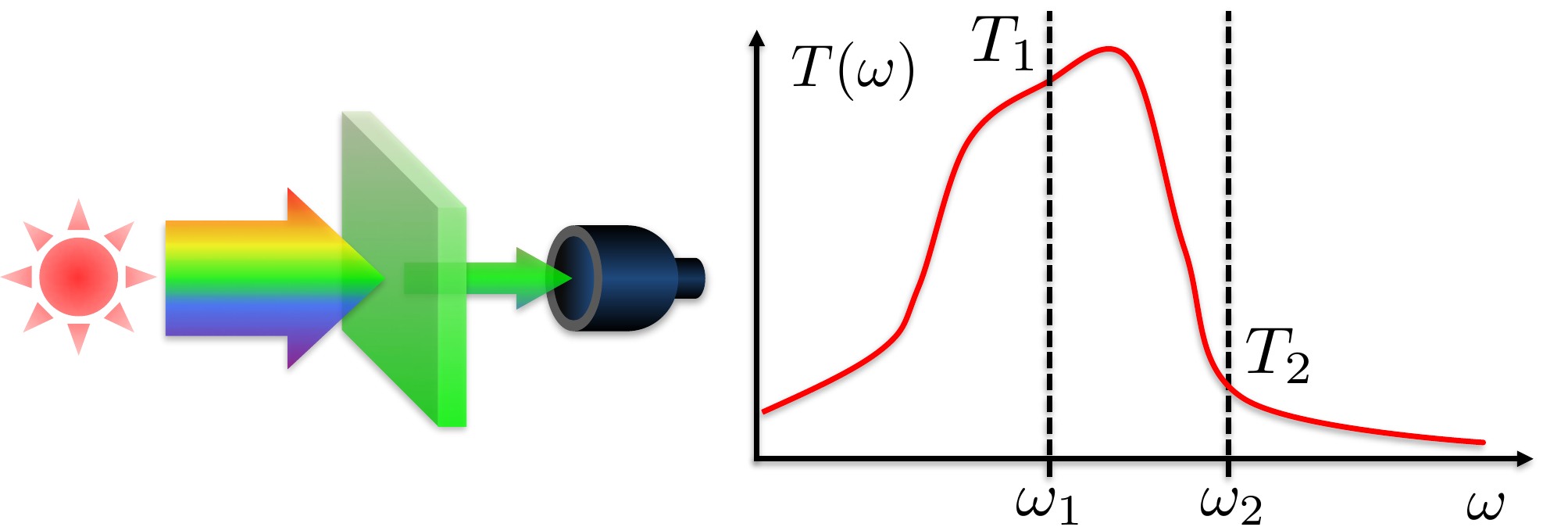}

\protect\caption{(Color online) Demonstration for the effective temperature distribution
of the bath modes. A filter in front of a thermal source can be utilized
to modify the effective temperature distribution, and the effective
temperature for each bath mode can be measured by a photon detector.}
\label{fig-demo}
\end{figure}

Thermal equilibrium is an idealistic physics model, and non-thermal
baths widely exist in realistic world \cite{olsina_can_2014,alicki_non-equilibrium_2015,xu_polaron_2016}.
We should say the thermal state is a good enough physical model within
certain finite space and time. For example, a simple filter in front
of a thermal light source can be utilized to create a non-thermal
state (Fig.\,\ref{fig-demo}). The states and the energy distribution
of the EM field modes on the two sides are obviously different.

Thus, we extend our requirement for the bath to be a non-thermal one,
but we still assume : 1. The total bath state is a product state of
each bath mode, i.e., $\rho_{B}=\rho_{k_{1}}\otimes\rho_{k_{2}}\otimes\dots$
2. Each mode state $\rho_{k}$ has a thermal form with an effective
temperature $T_{k}=\beta_{k}^{-1}$, i.e., $\rho_{k}\propto\exp[-\beta_{k}\,\omega_{k}\hat{b}_{k}^{\dagger}\hat{b}_{k}]$.
That is, the non-thermal state of the bath we study here is
\begin{equation}
\rho_{B}=\prod_{k}\frac{1}{{\cal Z}_{k}}\exp[-\frac{\omega_{k}}{T_{k}}\hat{b}_{k}^{\dagger}\hat{b}_{k}].\label{eq:Bath-state}
\end{equation}
In this non-thermal state, the different bath modes have a distribution
of effective temperatures $\{T_{k}\}$, but do not need to have a
constant value $T$. 

In principle, we can measure the state of each bath mode by optical
tomography, and obtain a Wigner function description for each optical
mode. For example, for the EM field under the non-thermal state (\ref{eq:Bath-state}),
the expectation value of the Poynting vector is
\begin{align}
\langle\hat{\boldsymbol{{\cal P}}}\rangle & =\frac{1}{\mu_{0}}\langle\hat{\mathbf{E}}\times\hat{\mathbf{B}}\rangle=\sum_{\mathbf{k},\sigma}\frac{c\,\hat{\mathrm{e}}_{\mathbf{k}}}{V}\cdot\hbar\omega_{\mathbf{k}}\big[\langle\hat{b}_{\mathbf{k}\sigma}^{\dagger}\hat{b}_{\mathbf{k}\sigma}\rangle+\frac{1}{2}\big]\nonumber \\
 & =\sum_{\mathbf{k},\sigma}\frac{c\,\hat{\mathrm{e}}_{\mathbf{k}}}{V}\cdot\hbar\omega_{\mathbf{k}}\overline{n}_{\mathrm{p}}(\omega_{\mathbf{k}},T_{\mathbf{k}}),
\end{align}
 where $\hat{\mathrm{e}}_{\mathbf{k}}$ is the direction of the wave
vector $\mathbf{k}$. $\hat{\mathbf{E}}=-\partial_{t}\hat{\mathbf{A}}$,
$\hat{\mathbf{B}}=\nabla\times\hat{\mathbf{A}}$, and $\hat{\mathbf{A}}$
is the field vector, 
\begin{equation}
\hat{\mathbf{A}}=\sum_{\mathbf{k}\sigma}\hat{\mathrm{e}}_{\mathbf{k}\sigma}\big[\frac{\hbar}{2\varepsilon_{0}\omega_{\mathbf{k}}V}\big]^{\frac{1}{2}}(\hat{b}_{\mathbf{k}\sigma}e^{i\mathbf{k}\cdot\mathbf{r}-i\omega_{\mathbf{k}}t}+\mathbf{h.c.}).
\end{equation}
Here we denote $\hat{\mathrm{e}}_{\mathbf{k}\sigma}$ as the polarization
direction \cite{scully_quantum_1997,orszag_quantum_2000}. 

The Poynting vector $\langle\hat{\boldsymbol{{\cal P}}}\rangle$ describes
the total energy flow passing through a unit section in unit time.
Each summation term in $\langle\hat{\boldsymbol{{\cal P}}}\rangle$
is the photon flux of certain optical mode, which is just the quantity
measured by the photon detector. Notice that in the above Poynting
vector, it contains $\overline{n}_{\mathrm{p}}(\omega,T):=[\exp(\omega/T)-1]^{-1}$,
which is the Planck distribution. Therefore, once we measure the light
intensity accepted by the photon detector, we obtain the effective
temperature $T_{\mathbf{k}}$ for a certain optical mode (Fig.\,\ref{fig-demo}).

In the idealistic case of a thermal bath, the temperatures for each
bath mode all equal to a constant value $T_{\mathbf{k}}=T$. In more
realistic cases, this requirement do not need to be fulfilled, and
so we obtain a distribution $\{T_{\mathbf{k}}\}$. If the distribution
$\{T_{\mathbf{k}}\}$ is far from a constant one, we say this non-thermal
bath state is far from the thermal equilibrium.

\section{Dynamics of a three-level system in a non-thermal bath}

\subsection{Born-Markovian approximation \label{sub:Born-Markovian-approximation}}

\begin{figure}
\includegraphics[width=0.98\columnwidth]{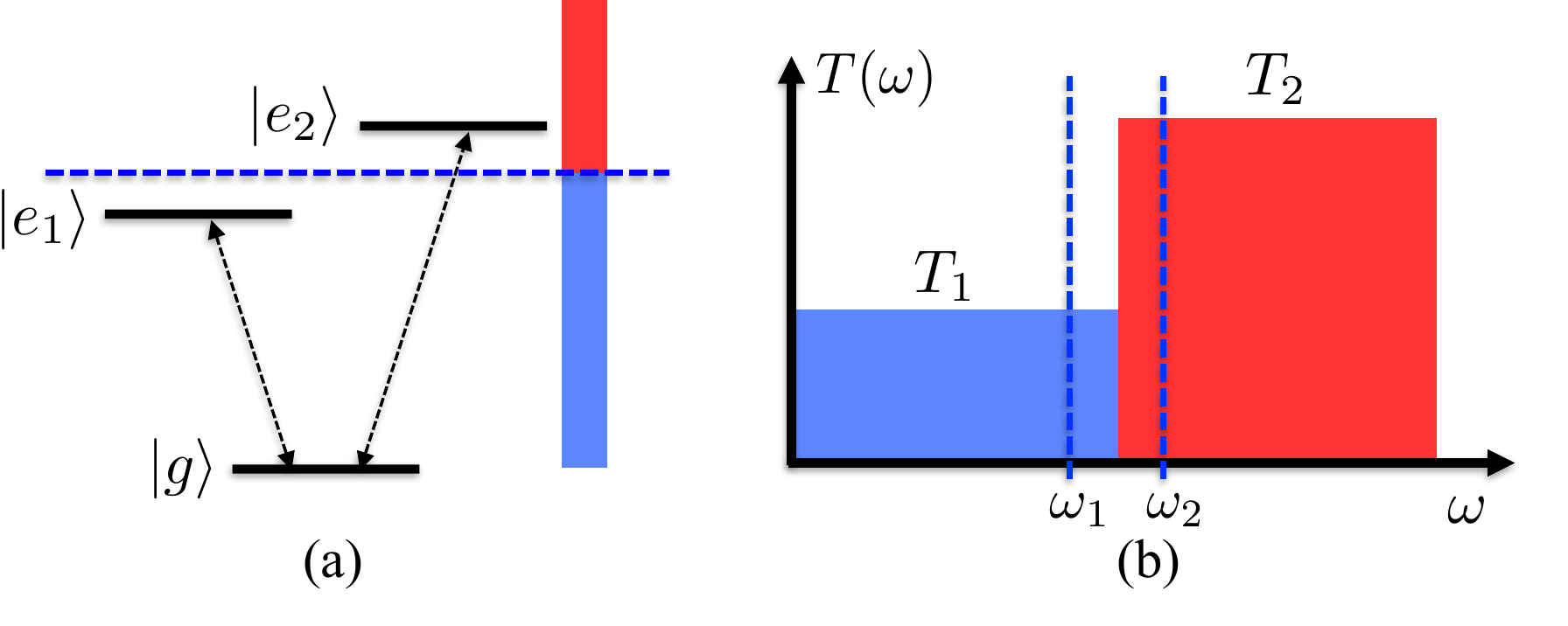}

\protect\caption{(Color online) Demonstration for (a) the three-level atom and its
transition structure, (b) the effective temperature distribution of
the bath modes. }
\label{fig-atom}
\end{figure}

Now we study the dynamics of a three-level system weakly coupled with
a non-thermal boson bath (Fig.\,\ref{fig-atom}). The Hamiltonian
of the system is
\begin{equation}
\hat{H}_{S}=\varepsilon_{1}|e_{1}\rangle\langle e_{1}|+\varepsilon_{2}|e_{2}\rangle\langle e_{2}|,
\end{equation}
and here we set the energy for the ground state $|g\rangle$ to be
$\varepsilon_{g}=0$. The interaction between the system and the bath
reads 
\begin{equation}
\hat{H}_{SB}=\sum_{n}\hat{\tau}_{n}^{+}\cdot\hat{B}_{n}+\hat{\tau}_{n}^{-}\cdot\hat{B}_{n}^{\dagger}
\end{equation}
where $\hat{B}_{n}=\sum_{k}\,g_{n,k}\hat{b}_{k}$ is the collective
operator of the bath, and $\hat{\tau}_{n}^{-}=|g\rangle\langle e_{n}|$,
$\hat{\tau}_{n}^{+}=|e_{n}\rangle\langle g|$ are the lowering and
raising operator of the 3-level system associate with the transition
between $|g\rangle$ and $|e_{n}\rangle$. Notice that here we have
made the rotating-wave approximation (RWA) and omitted all the double
creation/annihilation terms, because we consider that the interaction
strength is quite weak and the RWA still applies.

The bath is modeled as a collection of boson modes, $\hat{H}_{B}=\sum_{k}\omega_{k}\hat{b}_{k}^{\dagger}\hat{b}_{k}$,
as mentioned above. We need to derive a master equation to describe
the dynamics of this 3-level system, which comes from the iteration
of the von Neumann equation,
\begin{align}
\frac{d\rho_{S}}{dt}=\mathrm{Tr}_{B} & \Big\{-i[\rho_{SB}(t),\,\hat{H}_{SB}(t)]\nonumber \\
 & -\int_{0}^{t}ds\,\big[[\rho_{SB}(s),\,\hat{H}_{SB}(s)],\,\hat{H}_{SB}(t)\big]\Big\}.\label{eq:von-0}
\end{align}
This is an exact equation in the interaction picture of $\hat{H}_{S}+\hat{H}_{B}$,
but still not easy for practical calculations. 

We still need some assumptions to simplify the above equation. The
first one is the Born approximation, i.e., during the evolution, the
total state of the open system and the bath is $\rho_{SB}(t)\simeq\rho_{S}(t)\otimes\rho_{B}(0)$.
Namely, we consider the bath is so large that the system almost cannot
change bath state. Besides, the relaxation time of the bath $\tau_{B}$
is usually much shorter than the decay time of the open system $\tau_{S}$,
so the bath could ``refresh'' to its original state quickly before
the open system evolves \cite{breuer_theory_2002}. Here, although
the bath is a non-thermal state {[}Eq.\,(\ref{eq:Bath-state}){]},
we still assume the Born approximation is valid, so the above equation
becomes
\begin{align}
\frac{d\rho_{S}}{dt} & \simeq-\mathrm{Tr}_{B}\int_{0}^{t}ds\big[[\rho_{S}(s)\otimes\rho_{B},\hat{H}_{SB}(s)],\hat{H}_{SB}(t)\big]\label{eq:TC}\\
 & =-\mathrm{Tr}_{B}\int_{0}^{t}ds'\big[[\rho_{S}(t-s')\otimes\rho_{B},\hat{H}_{SB}(t-s')],\hat{H}_{SB}(t)\big].\nonumber 
\end{align}
Notice that the first term of Eq.\,(\ref{eq:von-0}) vanishes, because
for the non-thermal state Eq.\,(\ref{eq:Bath-state}), we always
have $\langle\hat{B}_{n}(t)\rangle=\mathrm{Tr}_{B}[\rho_{B}\cdot\sum g_{n,k}\hat{a}_{k}\exp(-i\omega_{k}t)]=0$.

Now we obtain a differential-integral equation for $\rho_{S}(t)$.
It contains a time-nonlocal convolution term, which means, if we want
to know the changing rate of $\rho_{S}(t)$, we need to accumulate
all the information of $\rho_{S}(s')$ for $0\le s'\le t$, not only
the instantaneous value of $\rho_{S}$ at $t$. 

At this stage, we further assume that the convolution kernel in the
integral, which comes from the time correlation function of the bath
operators, decays so fast with $t-s'$ that only the accumulation
around $\rho(t-s'\simeq t)$ dominates in the integral. Thus, from
Eq.\,(\ref{eq:TC}) we obtain
\begin{equation}
\frac{d\rho_{S}}{dt}\simeq-\mathrm{Tr}_{B}\int_{0}^{t}ds'\big[[\rho_{S}(t)\otimes\rho_{B},\hat{H}_{SB}(t-s')],\hat{H}_{SB}(t)\big].\label{eq:Mark-1}
\end{equation}
This is a differential equation local in time, and here we call this
approximation \emph{Markov-1}. If we use the cumulant expansion method
of van Kampen, also known as the time-convolutionless (TCL) method,
to derive a non-Markovian master equation, Eq.\,(\ref{eq:Mark-1})
is just its lowest order (TCL-2) \cite{breuer_theory_2002}.

Usually we further extend the integral to infinity approximately,
because we assumed that the time correlation functions of the bath
approach to their steady values quite fast, and here we call it \emph{Markov-2}.
After the approximation Markov-2, we obtain
\begin{equation}
\frac{d\rho_{S}}{dt}\simeq-\mathrm{Tr}_{B}\int_{0}^{\infty}ds'\big[[\rho_{S}(t)\otimes\rho_{B},\hat{H}_{SB}(t-s')],\hat{H}_{SB}(t)\big].\label{eq:Mark-2}
\end{equation}
This is usually the starting point for the derivation of the Markovian
master equation. 

The Liouville operator obtained from Markov-2 {[}Eq.\,(\ref{eq:Mark-2}){]}
usually does not depend on time explicitly (in Schr\"odinger's picture)
\cite{li_long-term_2014,li_steady_2015}, which brings us great convenience
for calculation, but we should keep in mind that Markov-2 indeed ignored
the precise dynamical behaviour within the bath correlation time $\tau_{B}$.
That is, for the period $0\le t\lesssim\tau_{B}$, the approximation
\begin{equation}
\int_{0}^{t}ds'...\simeq\int_{0}^{\infty}ds'...\label{eq:int-Mark-2}
\end{equation}
is indeed not quite reliable, but we just do not care about the dynamics
within this short time.

\subsection{Coherent transition and secular approximation}

Starting from Markov-2 {[}Eq.\,(\ref{eq:Mark-2}){]}, we can obtain
the following master equation in Schr\"odinger's picture (the details
of the derivation is shown in Appendix \ref{sec:Deri-MasEq}, see
also Ref.\,\cite{li_steady_2015}),
\begin{align}
\dot{\rho}_{S}= & i[\rho_{S},\hat{H}_{S}+\hat{H}_{\mathrm{c}}]\nonumber \\
 & +\sum_{m,n=1}^{2}D_{mn}^{+}\big(\hat{\tau}_{m}^{+}\rho_{S}\hat{\tau}_{n}^{-}-\frac{1}{2}\{\hat{\tau}_{n}^{-}\hat{\tau}_{m}^{+},\rho_{S}\}_{+}\big)\nonumber \\
 & +D_{mn}^{-}\big(\hat{\tau}_{m}^{-}\rho_{S}\hat{\tau}_{n}^{+}-\frac{1}{2}\{\hat{\tau}_{n}^{+}\hat{\tau}_{m}^{-},\rho_{S}\}_{+}\big),\label{eq:ME-mark}
\end{align}
where we denote the decay rates as $D_{mn}^{\pm}:=[\Gamma_{mn}^{\pm}(\varepsilon_{m})+\Gamma_{mn}^{\pm}(\varepsilon_{n})]/2$,
and 
\begin{align}
\Gamma_{mn}^{+}(\omega):= & J_{nm}(\omega)\tilde{n}(\omega),\nonumber \\
\Gamma_{mn}^{-}(\omega):= & J_{mn}(\omega)[\tilde{n}(\omega)+1].\label{eq:gamma(w)}
\end{align}
$J_{mn}(\omega)$ is the coupling spectrum, which is defined by 
\begin{equation}
J_{mn}(\omega):=2\pi\sum_{k}g_{m,k}^{*}g_{n,k}\delta(\omega-\omega_{k})=[J_{nm}(\omega)]^{*}.\label{eq:spectrum}
\end{equation}
 $\tilde{n}(\omega)$ is the photon number distribution obtained from
$\tilde{n}(\omega_{k})=\mathrm{Tr}[\rho_{B}\cdot\hat{a}_{k}^{\dagger}\hat{a}_{k}]$.
For a non-thermal state like Eq.\,(\ref{eq:Bath-state}), we have
\begin{equation}
\tilde{n}(\omega)=\cfrac{1}{\exp\frac{\omega}{T(\omega)}-1},
\end{equation}
where $T(\omega)$ is the effective temperature distribution. If $T(\omega)$
is a constant distribution $T(\omega)=T$, $\tilde{n}(\omega)$ could
return to the Planck distribution $\overline{n}_{\mathrm{p}}(\omega,T)$
for the thermal bath. 

Notice that except the dissipation part in the master equation, we
also have a correction in the unitary term \cite{li_steady_2015},
\begin{align}
\hat{H}_{\mathrm{c}}:=\sum_{m,n=1}^{2} & \frac{1}{4i}[\Gamma_{mn}^{+}(\varepsilon_{m})-\Gamma_{mn}^{+}(\varepsilon_{n})]\cdot\hat{\tau}_{n}^{-}\hat{\tau}_{m}^{+}\nonumber \\
 & +\frac{1}{4i}[\Gamma_{mn}^{-}(\varepsilon_{m})-\Gamma_{mn}^{-}(\varepsilon_{n})]\cdot\hat{\tau}_{n}^{+}\hat{\tau}_{m}^{-},\label{eq:H_c}
\end{align}
and $\hat{H}_{\mathrm{c}}$ is not from the principal integral. Here
we omitted the principal integrals in the above master equation.

From the definition of $J_{mn}(\omega)$ {[}Eq.\,(\ref{eq:spectrum}){]},
we directly obtain the property $J_{mn}(\omega)=[J_{nm}(\omega)]^{*}$,
and this guarantees $\hat{H}_{\mathrm{c}}=\hat{H}_{\mathrm{c}}^{\dagger}$.
Here we have both individual spectrums $J_{nn}(\omega)$ and cross
spectrums $J_{mn}(\omega)$ for $m\neq n$ \cite{wei_time_1994,li_steady_2015}.
The individual spectrum $J_{nn}(\omega)$ corresponds to the decay/excitation
behavior of each transition path itself, while the cross spectrum
$J_{mn}(\omega)$ $(m\neq n)$ corresponds to the coherent interference
between different transition paths. This effect is often called the
Fano-Agarwal interference \cite{agarwal_quantum_1974}. It can be
proved that we always have $\left|J_{mn}(\omega)\right|^{2}\le J_{mm}(\omega)\cdot J_{nn}(\omega)$
\cite{li_steady_2015}. Thus we define a coherence strength $p(\omega)$,
so that $J_{12}(\omega)=p(\omega)\cdot\sqrt{J_{11}(\omega)J_{22}(\omega)}$,
to measure the interference strength of the coherent transition. When
$\left|p(\omega)\right|=1$, we have the maximum interference effect,
while $p(\omega)=0$ indicates there is no interference between the
two transitions. For a specific physics model, this coherence strength
$p(\omega)$ can be calculated explicitly from the system-bath interaction.

It should be notice that, in the interaction picture, these coherent
transition terms have time-dependent coefficients $\exp[\pm i\Delta_{mn}t]$
when the two levels are not degenerated $\Delta_{mn}:=\varepsilon_{m}-\varepsilon_{n}\neq0$.
Thus, these terms are often dropped by the secular approximation in
order to directly get a Lindblad form with explicitly positive decay
rates \cite{breuer_theory_2002,wichterich_modeling_2007,creatore_efficient_2013}.
However, this secular approximation indeed implies that the precise
evolution details within the time scale $\tau_{\Delta}\simeq\hbar/\Delta_{mn}$
is omitted. When $\Delta_{mn}$ is a small value, $\tau_{\Delta}$
is comparable with the system decay time $\tau_{S}$, and thus this
is not a good enough approximation. Besides, from the above discussion
of the cross spectrum $J_{mn}(\omega)$, we see that these coherent
transition terms do have a clear physical meaning, and they play an
essential role in some physical effects like coherent population trapping
and dark state, which has been well known both theoretically and experimentally
\cite{agarwal_quantum_1974,zhu_quantum-mechanical_1995,scully_quantum_1997}.
Moreover, it was also found that these coherent transitions are closely
connected with the non-equilibrium flux in some transport systems,
and if they are omitted, we will obtain some unphysical results \cite{wichterich_modeling_2007,li_long-term_2014,li_steady_2015}.
Therefore, the secular approximation should not be made at this stage.

In sum, in the derivation of the master equation (\ref{eq:ME-mark}),
we used the Born approximation, Markov-1 and Markov-2. Usually when
we talk Markovian approximation in literature, we mean Markov-1 and
Markov-2 together. Now we are going to show that for the non-thermal
bath case, we need to separate them apart, and indeed Markov-2 is
not a good enough approximation. More importantly, we should keep
in mind that the name ``Markovian approximation'' never promises
us to give a legitimate Markovian dynamical behavior, as we will show
below.

\subsection{The problem of negative probability}

Now we can use the master equation (\ref{eq:ME-mark}) to study the
dynamics of the 3-level system. The influence of the non-thermal bath
state is contained in the modified photon number distribution $\tilde{n}(\omega)$.
For simplicity, we consider a distribution as follows {[}Fig.\,\ref{fig-atom}(b){]},
\begin{equation}
T(\omega)=\begin{cases}
T_{1}, & 0<\omega\le\overline{\varepsilon},\\
T_{2}, & \omega>\overline{\varepsilon},
\end{cases}\label{eq:temp-dist}
\end{equation}
where $\overline{\varepsilon}:=(\varepsilon_{1}+\varepsilon_{2})/2$
just lies in the middle between the two energy gap of the 3-level
system. 

\begin{figure}
\includegraphics[width=1\columnwidth]{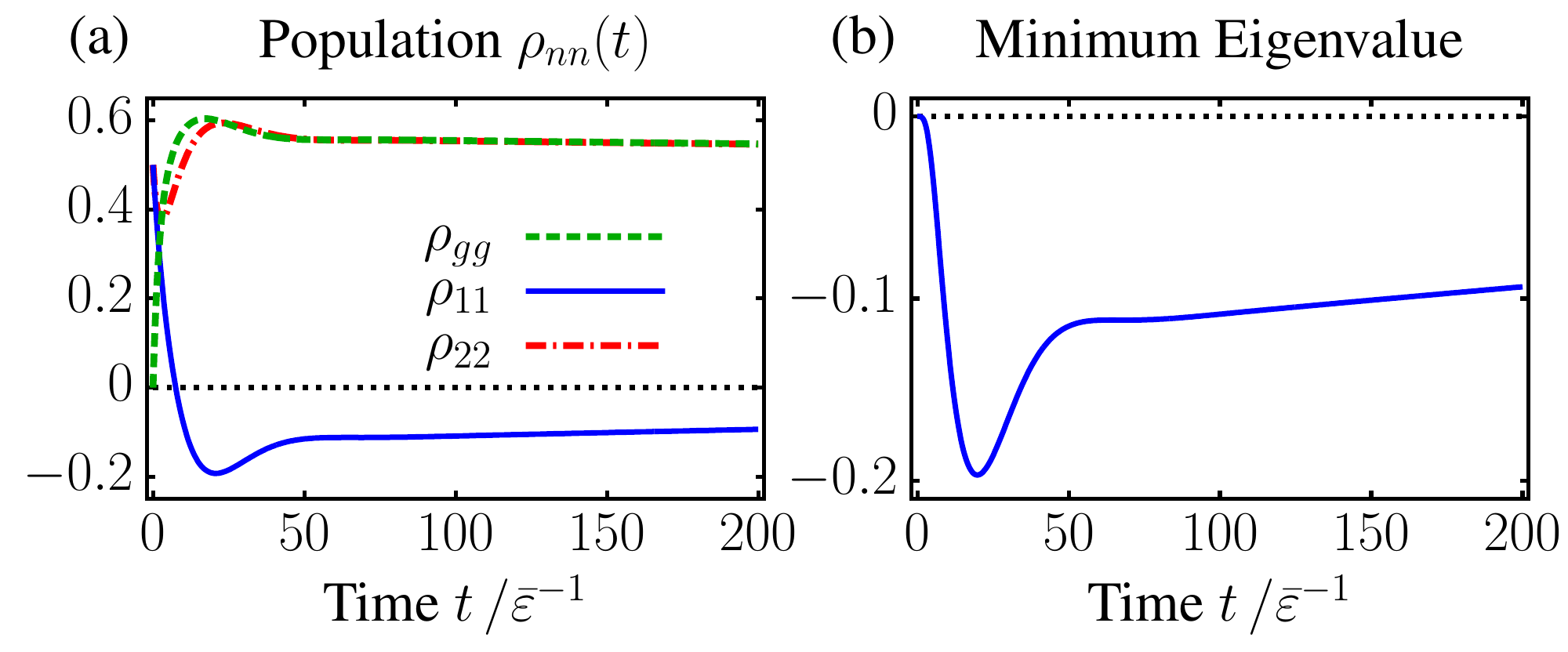}

\protect\caption{(Color online) Time-dependent evolution of $\rho_{S}(t)$ in a non-thermal
bath calculated from the master equation (\ref{eq:ME-mark}). (a)
The populations $\rho_{nn}(t)$ and (b) The smallest eigenvalue of
$\rho_{S}(t)$. $\rho_{11}(t):=\langle e_{1}|\rho_{S}(t)|e_{1}\rangle$
become negative. Here we set $\overline{\varepsilon}:=(\varepsilon_{1}+\varepsilon_{2})/2\equiv1$
as the unit, and $\varepsilon_{1}=0.95$, $\varepsilon_{2}=1.05$.
The interference strength is $p=1$. The decay rate is $J_{mn}(\varepsilon_{1,2})\equiv\gamma=0.0005$
for $m,n=1,2$. The two effective temperatures are $T_{1}=1$, $T_{2}=500$,
which is very far from the thermal equilibrium.}
\label{fig-Mark-2T}
\end{figure}

Due to the change of the photon number distribution, the decay rates
in the master equation does not satisfy the detailed balance or micro-reversibility
\cite{bergmann_new_1955,cai_entropy_2014}. Intuitively, if we do
not consider the coherent transition induced by the cross coupling
spectrum, we can regard the two excited state $|e_{1}\rangle$ and
$|e_{2}\rangle$ as two independent systems immersed in two thermal
baths with temperatures $T_{1}$ and $T_{2}$ respectively. 

We calculate the time-dependent evolution of the 3-level system in
a non-thermal bath according to the master equation (\ref{eq:ME-mark}),
and we show the result in Fig.\,\ref{fig-Mark-2T}. Strikingly, if
the evolutions start from the initial state $|\psi_{0}\rangle=(|e_{1}\rangle+|e_{2}\rangle)/\sqrt{2}$,
the population $\rho_{22}(t):=\langle e_{2}|\rho_{S}(t)|e_{2}\rangle$
on the energy level $|e_{2}\rangle$ becomes negative, and this negative
probability lasts for a very long time (Fig.\,\ref{fig-Mark-2T})
\cite{creatore_efficient_2013,dorfman_photosynthetic_2013}.

\begin{figure}
\includegraphics[width=1\columnwidth]{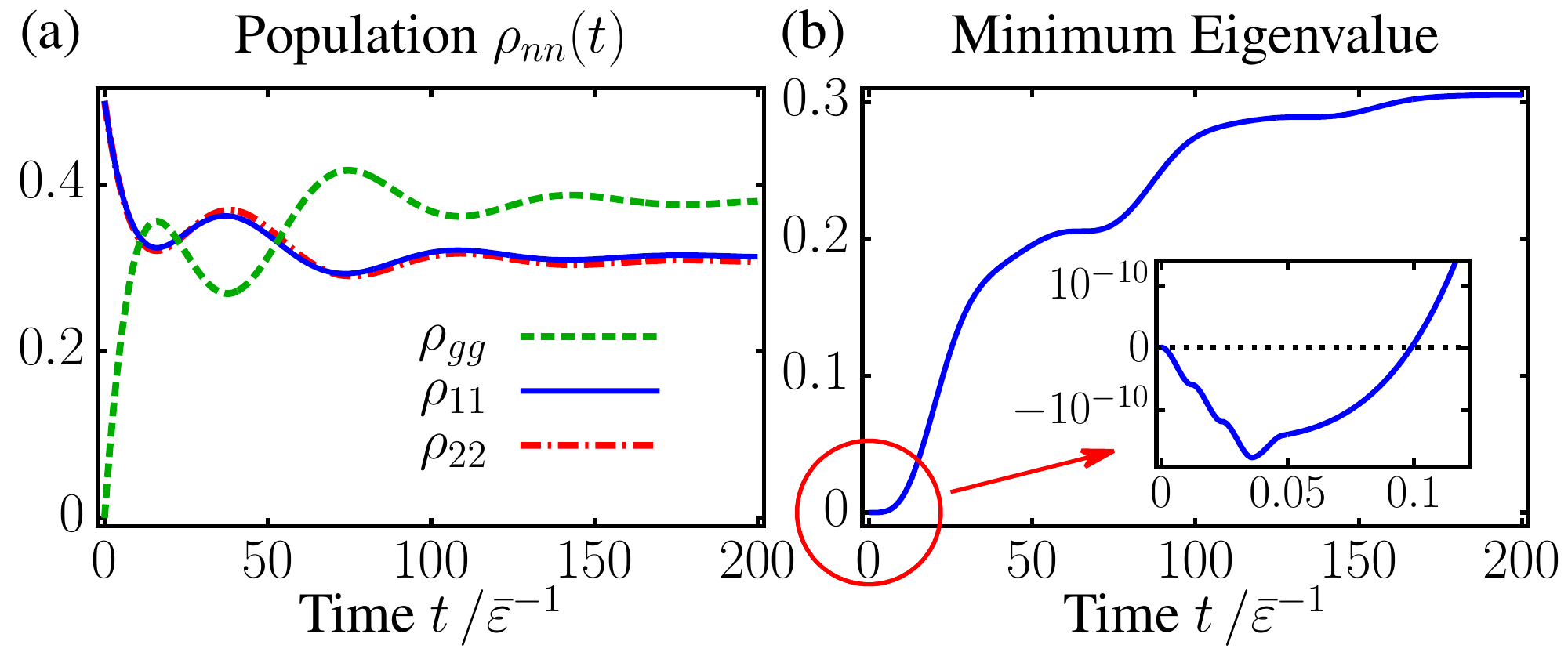}

\protect\caption{(Color online) Time-dependent evolution of $\rho_{S}(t)$ in a thermal
bath calculated from the master equation (\ref{eq:ME-mark}). (a)
The populations $\rho_{nn}(t)$ (b) The smallest eigenvalue of $\rho_{S}(t)$.
The populations in each energy levels are positive, but the minimum
eigenvalue of $\rho_{S}(t)$ becomes a small negative value immediately
after the evolution begins, and lasts for a very short time. Here
we set $\overline{\varepsilon}:=(\varepsilon_{1}+\varepsilon_{2})/2\equiv1$
as the unit, and $\varepsilon_{1}=0.95$, $\varepsilon_{2}=1.05$.
The interference strength is $p=1$. The decay rate is $J_{mn}(\varepsilon_{1,2})\equiv\gamma=0.01$
for $m,n=1,2$. The temperature of the thermal bath is $T=5$.}
\label{fig-Mark-1T}
\end{figure}

The diagonal terms $\rho_{nn}(t)$ of the density matrix means the
probabilities that the system stays in this state $|n\rangle$ at
time $t$, and they should never be negative. The density matrix must
be semi-positive at any time. Indeed this negative probability problem
also exists when the bath is a thermal state. We show the evolution
of the 3-level system when the bath is a thermal one (Fig.\,\ref{fig-Mark-1T}).
In this case, the populations in each eigen energy level are always
positive during the evolution starting from $|\psi_{0}\rangle=(|e_{1}\rangle+|e_{2}\rangle)/\sqrt{2}$
as before. But if we further check the minimum eigenvalue of $\rho_{S}(t)$,
we find that the smallest eigenvalue of $\rho_{S}(t)$ becomes a negative
value right after the evolution begins, and quickly turns to be positive
after a very short time. That means, $\rho_{S}(t)$ also has a negative
probability in a certain basis even when the bath is a thermal state.

For the thermal bath case, this negative probability problem is not
too serious, since the negative value is quite small, and it only
lasts for a very short time (Fig.\,\ref{fig-Mark-1T}). In the sense
of the coarse-graining idea of Markovian approximation, this inaccuracy
is negligible \cite{suarez_memory_1992}. But this problem is intolerable
when the bath is a non-thermal state far from equilibrium, because
the negative value is significantly large and lasts for too long time
(Fig.\,\ref{fig-Mark-2T}).

To find the reason for this negative probability problem, we need
to check the positivity condition of the master equation (\ref{eq:ME-mark}).
It was proved that for a master equation like
\begin{align}
\dot{\rho} & ={\cal L}[\rho]\nonumber \\
 & =i[\rho,\hat{H}]+\sum_{m,n}\kappa_{mn}\big(\hat{L}_{m}\rho\hat{L}_{n}^{\dagger}-\frac{1}{2}\{\hat{L}_{n}^{\dagger}\hat{L}_{m},\,\rho\}_{+}\big),
\end{align}
starting from any legitimate density matrix $\rho_{0}$, $\rho(t)$
is always semi-positive at any time $t$, if and only if the coefficients
$\kappa_{mn}$ form a non-negative Hermitian matrix \cite{gorini_completely_1976,lindblad_generators_1976}.
This equation is called the Lindblad form or GKSL form (Gorini-Kossakowski-Sudarshan-Lindblad).
Here $\hat{H}$ is not required to be the self Hamiltonian of the
system, and it is only required that $\hat{H}$ should be a Hermitian
operator.

The master equation (\ref{eq:ME-mark}) has already been written in
a Lindblad form, but we need to check the positivity of the decay
rate matrix $\boldsymbol{D}$ for the Lindblad operators $\{\hat{\tau}_{1}^{+},\,\hat{\tau}_{2}^{+},\,\hat{\tau}_{1}^{-},\,\hat{\tau}_{2}^{-}\}$,
i.e., 
\begin{align}
\boldsymbol{D} & =\left[\begin{array}{cc}
\boldsymbol{D}^{+} & \mathbf{0}\\
\mathbf{0} & \boldsymbol{D}^{-}
\end{array}\right],\\
\boldsymbol{D}^{\pm} & =\left[\begin{array}{cc}
\Gamma_{11}^{\pm}(\varepsilon_{1}) & \cfrac{1}{2}\big[\Gamma_{12}^{\pm}(\varepsilon_{1})+\Gamma_{12}^{\pm}(\varepsilon_{2})\big]\\
\cfrac{1}{2}\big[\Gamma_{12}^{\pm}(\varepsilon_{1})+\Gamma_{12}^{\pm}(\varepsilon_{2})\big] & \Gamma_{22}^{\pm}(\varepsilon_{2})
\end{array}\right],\nonumber 
\end{align}
 and the definition of $\Gamma_{mn}^{\pm}(\omega)$ follows from Eq.\,(\ref{eq:gamma(w)}).
To prove the positivity of $\boldsymbol{D}$, it suffices to prove
the positivity of the two block matrices $\boldsymbol{D}^{\pm}$.
It is obvious we already have $\Gamma_{nn}^{\pm}(\omega)\ge0$, thus
we just need to check whether $\det\boldsymbol{D}^{\pm}\ge0$, i.e.,
\begin{equation}
\det\boldsymbol{D}^{\pm}=\Gamma_{11}^{\pm}(\varepsilon_{1})\Gamma_{22}^{\pm}(\varepsilon_{2})-\frac{1}{4}\left|\Gamma_{12}^{\pm}(\varepsilon_{1})+\Gamma_{12}^{\pm}(\varepsilon_{2})\right|^{2}.
\end{equation}
 In the above two examples, we assumed $J_{nn}(\varepsilon_{n})=\gamma_{n}$,
and $J_{12}(\varepsilon_{1})=J_{12}(\varepsilon_{2})=p\cdot\sqrt{\gamma_{1}\gamma_{2}}$,
$0\le p\le1$, thus the determinants are
\begin{align}
\det\boldsymbol{D}^{+}= & (1-p^{2})\gamma_{1}\tilde{n}(\varepsilon_{1})\cdot\gamma_{2}\tilde{n}(\varepsilon_{2})\nonumber \\
 & \quad-\frac{p^{2}}{4}\cdot\gamma_{1}\gamma_{2}[\tilde{n}(\varepsilon_{1})-\tilde{n}(\varepsilon_{2})]^{2},\nonumber \\
\det\boldsymbol{D}^{-}= & (1-p^{2})\gamma_{1}[\tilde{n}(\varepsilon_{1})+1]\cdot\gamma_{2}[\tilde{n}(\varepsilon_{2})+1]\nonumber \\
 & \quad-\frac{p^{2}}{4}\cdot\gamma_{1}\gamma_{2}[\tilde{n}(\varepsilon_{1})-\tilde{n}(\varepsilon_{2})]^{2}.\label{eq:Det}
\end{align}
Therefore, if the coherent transition achieves the maximum interference
$p=1$, we always have $\det\boldsymbol{D}^{\pm}\le0$, and the equality
holds if and only if $\tilde{n}(\varepsilon_{1})=\tilde{n}(\varepsilon_{2})$.
That means, even for the case of a thermal bath $\tilde{n}(\varepsilon_{n})=\overline{n}_{\mathrm{p}}(\varepsilon_{n},T)$,
the decay matrix $\boldsymbol{D}$ always contains negative eigenvalues
unless $\varepsilon_{1}=\varepsilon_{2}$ or $T=0$. 

The negative eigenvalue of the decay matrix $\boldsymbol{D}$ implies
some modes of the system have negative decay rates. This is also why
the problem of negative probability appears in the master equation
(\ref{eq:ME-mark}) after the approximation Markov-2. For a thermal
bath state, the difference of $\overline{n}_{\mathrm{p}}(\varepsilon_{1},T)$
and $\overline{n}_{\mathrm{p}}(\varepsilon_{2},T)$ are quite small
when $\Delta_{12}/T\ll1$, thus $\det\boldsymbol{D}^{\pm}$ are very
small although negative, and so do the negative eigenvalues of $\boldsymbol{D}$.
But when the bath is a non-thermal state, the difference of $\tilde{n}(\varepsilon_{1})$
and $\tilde{n}(\varepsilon_{2})$ can be very large, and that gives
rise to very large negative diverging rate. Therefore, the problem
of negative probability is much more serious in a non-thermal bath
as mentioned in Fig.\,\ref{fig-Mark-2T}.

From the determinants Eq.\,(\ref{eq:Det}) we also notice that the
decay rate matrix $\boldsymbol{D}$ is always positive if the coherence
strength is lower than a upper bound, 
\begin{equation}
p\le p_{\mathrm{c}}:=\frac{4\tilde{n}(\varepsilon_{1})\tilde{n}(\varepsilon_{2})}{[\tilde{n}(\varepsilon_{1})+\tilde{n}(\varepsilon_{2})]^{2}}\le1.\label{eq:p_c}
\end{equation}
And the larger the difference $|\tilde{n}(\varepsilon_{1})-\tilde{n}(\varepsilon_{2})|$
is, the smaller $p_{c}$ we obtain. This bound is also the limit for
the validity of the Markovian master equation with coherent transition,
namely, if the strength of the coherent transition is greater than
this bound $p_{\mathrm{c}}$, it is no longer possible to describe
the system dynamics by a homogenous Markovian master equation like
Eq.\,(\ref{eq:ME-mark}).

\section{Non-Markovian Dynamics}

We have seen that if the interference strength of the coherent transition
is greater than the upper limit $p_{\mathrm{c}}$ {[}Eq.\,(\ref{eq:p_c}){]},
the positivity of the decay rate matrix $\boldsymbol{D}$ in the homogeneous
Markovian master equation (\ref{eq:ME-mark}) is broken down, and
that brings in the problem of negative probability. This problem is
intolerable when the bath state is far from thermal equilibrium because
the negative probability has a significantly large value and lasts
for a quite long time (Fig.\,\ref{fig-Mark-2T}).

However, the interference strength $p(\omega)$ of the coherent transition
is determined by the properties of the physical system itself, and
there is no physical law forbidding it to achieve its maximum value.
Thus, now we need to answer, if a system with maximum coherent transition
is immersed in a non-thermal bath, how to resolve this negative probability
problem. Here we show that this problem can be cured by considering
the time-dependence of the decay rates, no matter whether the bath
is thermal or non-thermal. And we will see that this gives rise to
non-Markovian dynamics.

\subsection{Correlation time}

First, we release the approximation Markov-2 {[}Eq.\,(\ref{eq:Mark-2}){]},
and derive another time-dependent master equation based on Markov-1
{[}Eq.\,(\ref{eq:Mark-1}){]}, which reads (see derivations in Appendix
\ref{sec:Deri-MasEq}),
\begin{align}
\dot{\rho}_{S}= & i[\rho_{S},\hat{H}_{S}+\hat{H}_{\mathrm{c}}(t)]\nonumber \\
 & +\sum_{m,n=1}^{2}D_{mn}^{+}(t)\big(\hat{\tau}_{m}^{+}\rho_{S}\hat{\tau}_{n}^{-}-\frac{1}{2}\{\hat{\tau}_{n}^{-}\hat{\tau}_{m}^{+},\rho_{S}\}_{+}\big)\nonumber \\
 & +D_{mn}^{-}(t)\big(\hat{\tau}_{m}^{-}\rho_{S}\hat{\tau}_{n}^{+}-\frac{1}{2}\{\hat{\tau}_{n}^{+}\hat{\tau}_{m}^{-},\rho_{S}\}_{+}\big).\label{eq:ME-non-Mark}
\end{align}
Comparing with the homogeneous Markovian master equation (\ref{eq:ME-mark}),
the decay rates become time-dependent now, i.e., $D_{mn}^{\pm}(t):=[\Gamma_{mn}^{\pm}(\varepsilon_{m},t)+\Gamma_{mn}^{\pm}(\varepsilon_{n},t)]/2$,
and 
\begin{align}
\Gamma_{mn}^{+}(\omega,t): & =\Re\mathrm{e}\int_{0}^{t}ds\int_{0}^{\infty}\frac{d\nu}{2\pi}J_{nm}(\nu)\tilde{n}(\nu)e^{i(\nu-\omega)s}\label{eq:Gam(w,t)}\\
\Gamma_{mn}^{-}(\omega,t): & =\Re\mathrm{e}\int_{0}^{t}ds\int_{0}^{\infty}\frac{d\nu}{2\pi}J_{mn}(\nu)[\tilde{n}(\nu)+1]e^{i(\omega-\nu)s}.\nonumber 
\end{align}
Notice that $\Gamma_{mn}^{-}(\omega,t)=\Gamma_{nm}^{+}(\omega,t)+\Gamma_{mn}^{0}(\omega,t)$,
where 
\begin{equation}
\Gamma_{mn}^{0}(\omega,t):=\Re\mathrm{e}\int_{0}^{t}ds\int_{0}^{\infty}\frac{d\nu}{2\pi}\,J_{mn}(\nu)e^{i(\nu-\omega)s}
\end{equation}
does not depend on the bath state. The unitary correction term $\hat{H}_{\mathrm{c}}(t)$
is the same as Eq.\,(\ref{eq:H_c}) except the corresponding changes
of $\Gamma_{mn}^{\pm}(\varepsilon_{i},t)$. When $t\rightarrow\infty$,
$\Gamma_{mn}^{\pm}(\omega,t)$ could return to Eq.\,(\ref{eq:gamma(w)})
exactly. The imaginary part of the above integrals contributes to
a time-dependent Lamb shift which we do not consider here.

\begin{figure}
\includegraphics[width=1\columnwidth]{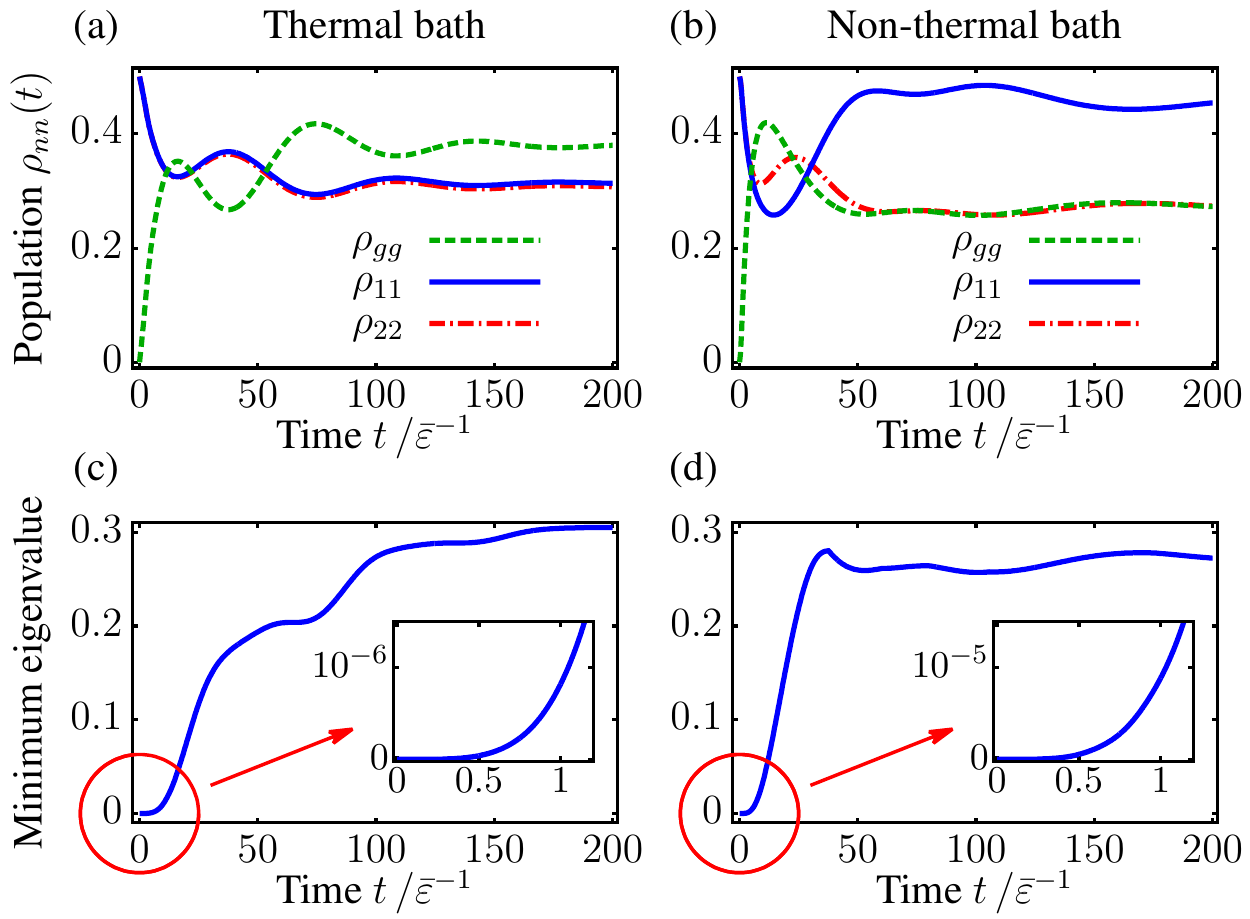}

\protect\caption{(Color online) (a, b) Evolution of the populations $\rho_{nn}(t)$
in thermal/non-thermal bath calculated from the time-dependent master
equation (\ref{eq:ME-non-Mark}). (c, d) The minimum eigenvalue of
$\rho_{S}(t)$ bath also keeps non-negative in thermal/non-thermal.
The coupling spectrums are $J_{mn}(\omega)=\lambda\omega\cdot\Theta(\Omega_{\mathrm{c}}-\omega)$.
Here we set $\overline{\varepsilon}:=(\varepsilon_{1}+\varepsilon_{2})/2\equiv1$
as the unit, and $\varepsilon_{1}=0.95$, $\varepsilon_{2}=1.05$.
The interference strength is $p=1$. We choose the cutoff as $\Omega_{\mathrm{c}}=20$.
For the thermal case (a, c), we set $T=5$ and $\lambda=0.01$ (see
also Fig.\,\ref{fig-Mark-1T}). For the non-thermal case (b, d) we
set $T_{1}=1$, $T_{2}=500$ and $\lambda=0.0005$ (see also Fig.\,\ref{fig-Mark-2T}). }
\label{fig-non-Mark}
\end{figure}

When the time dependence of the decay rates are considered, we need
to choose a specific coupling spectrum $J_{mn}(\omega)$ {[}Eq.\,(\ref{eq:spectrum}){]}.
Here we use the linear spectrum $J_{mn}(\omega)=\lambda_{mn}\omega$
for $m,n=1,2$ (also known as the Ohmic spectrum), with a step function
$\Theta(\Omega_{c}-\omega)$ as the cutoff. When the cutoff $\Omega_{c}\rightarrow\infty$,
this linear coupling spectrum could lead to white noise, namely, the
noise spectrum of the bath tends to be flat \cite{breuer_theory_2002,clerk_introduction_2010,weiss_quantum_2012},
which fits the idea of Markovian approximation more closely. For simplicity,
we choose $\lambda_{mn}=\lambda$ for $m,n=1,2$, which also implies
the interference of the coherent transition achieves the maximum $p=1$.
Accordingly, $\Gamma_{mn}^{\pm}(\omega,t)$ equal to each other for
$m,n=1,2$, and we denote as $\Gamma_{mn}^{\pm}(\omega,t):=\Gamma^{\pm}(\omega,t)$.

We still start from the same initial state $|\psi_{0}\rangle=(|e_{1}\rangle+|e_{2}\rangle)/\sqrt{2}$
as before, and calculate the evolution of $\rho_{S}(t)$ by this time-dependent
master equation (\ref{eq:ME-non-Mark}) under the same parameters
as those in Figs.\,\ref{fig-Mark-2T}, \ref{fig-Mark-1T} correspondingly.
We find that, no matter whether the bath is thermal or non-thermal,
the populations $\rho_{nn}(t)$ are always positive, and so does the
minimum eigenvalue of $\rho_{S}(t)$ (Fig.\,\ref{fig-non-Mark}).
The previous negative probability problem is resolved.

This result implies that the application of Markov-2 is not appropriate,
which leads to the emergence of the negative probability problem as
the result. The reason is, when we do the Markovian approximations
as discussed in Sec.\,\ref{sub:Born-Markovian-approximation}, we
have assumed that the system decay time $\tau_{S}$ is much shorter
than the bath relaxation time $\tau_{B}$, but indeed this assumption
is still waiting for a self-consistency examination after we obtain
the master equation.

\begin{figure}
\includegraphics[width=1\columnwidth]{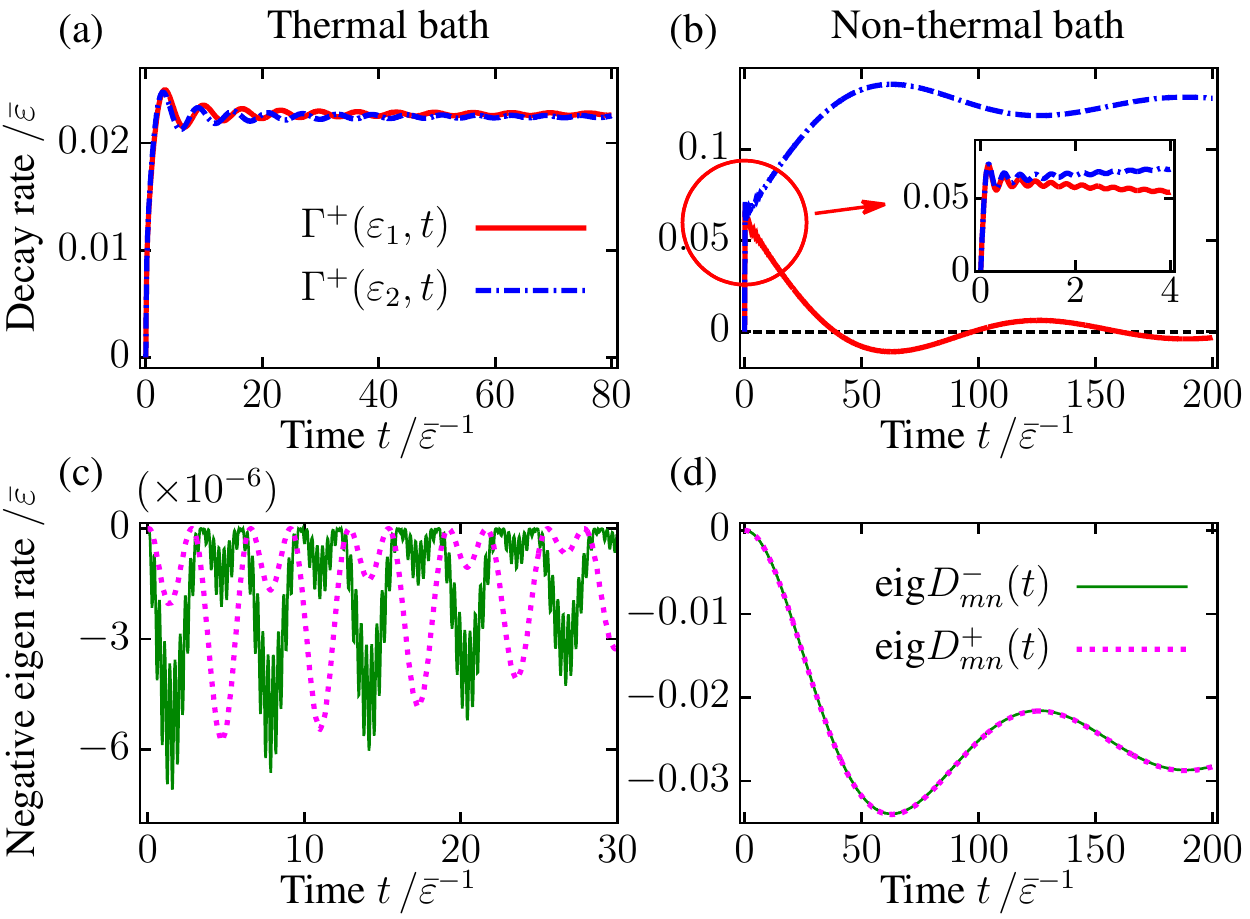}

\protect\caption{(Color online) (a, b) Time-dependent decay rates $\Gamma^{+}(\varepsilon_{1,2},t)$
in thermal/non-thermal bath. (c, d) Evolution of the negative eigenvalue
of the decay matrix $\boldsymbol{D}^{\pm}(t)$ in thermal/non-thermal
bath. The coupling spectrums are $J_{mn}(\omega)=\lambda\omega\cdot\Theta(\Omega_{\mathrm{c}}-\omega)$.
Here we set $\overline{\varepsilon}:=(\varepsilon_{1}+\varepsilon_{2})/2\equiv1$
as the unit, and $\varepsilon_{1}=0.95$, $\varepsilon_{2}=1.05$.
The interference strength is $p=1$. We choose the cutoff as $\Omega_{\mathrm{c}}=20$.
For the thermal case (a, c), we set $T=5$ and $\lambda=0.01$ (see
also Fig.\,\ref{fig-Mark-1T}). For the non-thermal case (b, d) we
set $T_{1}=1$, $T_{2}=500$ and $\lambda=0.0005$ (see also Fig.\,\ref{fig-Mark-2T}).
The two lines in (d) are almost overlapped.}
\label{fig-decay-rate}
\end{figure}

Now we do this consistency examination. We calculate the time dependence
of the decay rates $\Gamma^{+}(\varepsilon_{1,2},t)$ and show them
in Fig.\,\ref{fig-decay-rate}(a, b). For the thermal bath, $\Gamma^{+}(\varepsilon_{1,2},t)$
quickly reach their steady values after a very short time, and then
only have small oscillations around it. Therefore, it is a good enough
approximation to replace $\Gamma^{+}(\varepsilon_{1,2},t)$ by their
constant steady values. This is just what is done in the Markov-2
approximation.

We also see that this replacement by a constant is not good for $\Gamma^{+}(\varepsilon_{1,2},t)$
in the beginning short period. As the result, the dynamics calculated
from Markov-2 is not reliable within this period even for the case
of a thermal bath {[}see also discussion around Eq.\,(\ref{eq:int-Mark-2}){]}.
This is also why we have negative probability in the beginning period
shown in Fig.\,\ref{fig-Mark-1T}. In Markovian approximations, this
short time inaccuracy is omitted based on the idea of coarse-graining,
and we do notice that there is almost no difference between Fig.\,\ref{fig-Mark-1T}(a)
and Fig.\,\ref{fig-non-Mark}(a) even quantitatively.

For the case of a non-thermal bath far from equilibrium, the evolution
of $\Gamma^{+}(\varepsilon_{1,2},t)$ can be also divided into two
stages. However, after the first rapid relaxation stage, the decay
rates $\Gamma^{+}(\varepsilon_{1,2},t)$ still vary significantly
and slowly before they approach the steady values due to the non-thermal
temperature distribution. The decay rate $\Gamma^{+}(\varepsilon_{1},t)$
even becomes negative at some time. Obviously we cannot replace them
simply by a constant as the case of thermal bath.

We also notice that $\Gamma^{+}(\varepsilon_{2},t)$ tends to a large
value. This is because the frequency $\varepsilon_{2}$ is resonant
with a bath mode with a very high effective temperature $(T_{2}=500)$.
We should remember that the relaxation time $\tau_{S}$ of the open
system is characterized by $\Gamma^{\pm}(\varepsilon_{1,2},t)$. Thus,
the system relaxation time $\tau_{S}$ becomes much shorter. All these
facts imply that the previous basis ansatz for Markovian approximation,
$\tau_{B}\ll\tau_{S}$, no longer holds for a non-thermal bath far
from equilibrium, and the dynamics of the open system must be non-Markovian,
even if the coupling strength between the system and the bath is very
weak ($\lambda=0.0005$).

\subsection{Non-Markovianity}

Besides the comparison of the correlation time as above, we have some
more sufficient evidences to determine the non-Markovianity of the
open quantum system. The first simple non-Markovian feature is reflected
in the colored noise spectrum of the bath. The noise spectrum measured
by the detector is calculated from the Fourier transform of the symmetrized
time correlation function of the bath operators \cite{breuer_theory_2002,gardiner_quantum_2004,clerk_introduction_2010},
i.e.,
\begin{align}
\overline{S}_{mn}(\omega): & =\frac{1}{2}\int_{-\infty}^{\infty}ds\,\mathrm{Tr}\Big[\rho_{B}\cdot\{\hat{B}_{m}^{+}(t),\,\hat{B}_{n}^{-}(t+s)\}_{+}\Big]e^{i\omega s}\nonumber \\
 & =J_{mn}(\omega)[\tilde{n}(\omega)+\frac{1}{2}].
\end{align}
For the thermal bath case, we have $\tilde{n}(\omega)=\overline{n}_{\mathrm{p}}(\omega,T)$.
If we use the linear spectrum $J_{mn}(\omega)=\lambda_{mn}\omega$,
with the help of the expansion, 
\begin{equation}
\omega[\frac{1}{e^{\beta\omega}-1}+\frac{1}{2}]\simeq\frac{1}{\beta}+\frac{\beta\omega^{2}}{12}+...
\end{equation}
we obtain a flat noise spectrum $\overline{S}_{mn}(\omega)\simeq\lambda_{mn}T$
at high temperature limit, $\beta=T^{-1}\rightarrow0$, i.e., $\overline{S}_{mn}(\omega)$
is a constant and does not depend on $\omega$. This is just the white
noise spectrum which leads to Markovian dynamics \cite{breuer_theory_2002,gardiner_quantum_2004}.
However, if the bath has a non-trivial temperature distribution {[}e.g.,
like Eq.\,(\ref{eq:temp-dist}){]}, the noise spectrum would become
$\overline{S}_{mn}(\omega)\simeq\lambda_{mn}T(\omega)$, which is
a colored noise spectrum. That implies the emergence of non-Markovian
dynamics.

There are many different approaches to determine the non-Markovianity
of a quantum process \cite{rivas_quantum_2014}. Based on the divisibility
criterion of the dynamical map, it was proved that a master equation
like
\begin{align}
\dot{\rho} & ={\cal L}_{(t)}[\rho]\label{eq:Lindblad-t}\\
 & =i[\rho,\hat{H}(t)]+\sum_{m,n}\kappa_{mn}(t)\big(\hat{L}_{m}\rho\hat{L}_{n}^{\dagger}-\frac{1}{2}\{\hat{L}_{n}^{\dagger}\hat{L}_{m},\,\rho\}_{+}\big)\nonumber 
\end{align}
gives a quantum Markovian process, if and only if the decay matrix
$[\kappa_{mn}(t)]$ is semi-positive at any time $t$ \cite{chruscinski_markovianity_2012,rivas_quantum_2014}. 

Thus, we check the positivity of the decay matrix of the time-dependent
master equation (\ref{eq:ME-non-Mark}). When we have maximum interference
$p=1$ and $\Gamma_{mn}^{\pm}(\varepsilon_{1,2},t)=\Gamma^{\pm}(\varepsilon_{1,2},t)$
as above, we can prove that the determinants of the block matrices,
\begin{equation}
\boldsymbol{D}^{\pm}(t)=\left[\begin{array}{cc}
\Gamma^{\pm}(\varepsilon_{1},t) & \cfrac{1}{2}\big[\Gamma^{\pm}(\varepsilon_{1},t)+\Gamma^{\pm}(\varepsilon_{2},t)\big]\\
* & \Gamma^{\pm}(\varepsilon_{2},t)
\end{array}\right],
\end{equation}
 in the decay matrix $\boldsymbol{D}(t)$ are always negative, $\det\boldsymbol{D}^{\pm}(t)\le0$.
That means $\boldsymbol{D}(t)$ always have one positive and one negative
eigenvalue. We show the evolution of these two negative eigenvalues
in Fig.\,\ref{fig-decay-rate}(c, d) for both thermal/non-thermal
bath cases. These eigenvalues are negligibly small in the thermal
bath case, while significantly large in the non-thermal bath. For
the non-thermal bath case, even the diagonal terms of $\boldsymbol{D}(t)$
themselves, $\Gamma^{\pm}(\varepsilon_{1,2},t)$, could also become
negative {[}Fig.\,\ref{fig-decay-rate}(b){]}. Therefore, even when
there is no coherent transition, $p=0$, it is still necessary to
consider the non-Markovian effect in a non-thermal bath. As we mentioned
before, ``Markovian approximation'' did not promise to give a Markovian
process.

Therefore, we conclude that the dynamics of the open quantum system
shows typical non-Markovian feature even when the system-bath coupling
is very weak. It is worth noticing that in previous literatures non-Markovian
dynamics is usually resulted from the ultra-strong coupling between
the system and its environment \cite{hu_quantum_1992,breuer_theory_2002,ma_entanglement_2012,yang_master_2013,li_approach_2014,weiss_quantum_2012}.
Different from previous studies, the non-Markovianity in our study
roots from the coherent transition and the non-thermal property of
the bath. Especially, the non-Markovianity is greatly enhanced by
the non-thermality of the bath.

Here we need to emphasize that the above criterion for Markovian process
did not include how to determine whether the time-dependent master
equation (\ref{eq:Lindblad-t}) is complete positive, and this is
still an open question \cite{hall_complete_2008}. Indeed, if we keep
increasing the system-bath coupling strength $\lambda$ or the temperature
difference $T_{2}-T_{1}$ in the above examples, the time-dependent
master equation (\ref{eq:ME-non-Mark}) will also have the negative
problem. In that case, non-Markovian corrections of higher orders
are needed.

\section{Conclusion}

In this paper, we studied the dynamics of an open quantum system interacting
with a non-thermal bath. We find that when the bath state is far from
thermal equilibrium, the dynamics of the open quantum system becomes
non-Markovian. The noise spectrum of the non-thermal bath is not the
flat one as the white noise case. As a result, the correlation functions
of the non-thermal bath vary significantly for a long time during
relaxation. These behaviors are no longer consistent with the Markovian
master equation which is widely adopted in previous studies.

The coherent transition in the open quantum system could bring in
the negative probability problem. For the thermal bath case, this
problem is negligible in the sense of coarse-graining idea of Markovian
approximation. But it is intolerable in a non-thermal bath, where
the negative probability is large and exists for very long time. This
negative probability problem can be naturally resolved after we consider
non-Markovian corrections.

Different from previous studies, here even when the system is weakly
coupled with the bath, we still cannot find a coarse-grained Markovian
description for the open quantum system in a non-thermal bath. the
non-Markovianity in our study is resulted from the coherent transition
and the non-thermality of the bath. The non-Markovianity can be greatly
enhanced by the non-thermality of the bath. 

\emph{Acknowledgement - }S.-W. Li appreciates much for the helpful
discussions with H. Dong, A. Svidzinsky, and D. Wang in Texas A\&M
University. This study is supported by NSF Grant PHY-1241032 and Robert
A. Welch Foundation Award A-1261.

\begin{widetext}

\appendix

\section{Derivation of the Master equations\label{sec:Deri-MasEq}}

Here we show the detailed derivations for the master equation (\ref{eq:ME-mark})
and the time-dependent one Eq.\,(\ref{eq:ME-non-Mark}). First, through
the approximation Markov-2 {[}Eq.\,(\ref{eq:Mark-2}){]}, we have
the following equation in the interaction picture, 
\begin{align}
\dot{\rho}_{S} & =-\mathrm{Tr}_{B}\int_{0}^{\infty}ds\,[H_{SB}(t),[H_{SB}(t-s),\rho_{S}(t)\otimes\rho_{B}]]\nonumber \\
 & =\mathrm{Tr}_{B}\int_{0}^{\infty}ds\,\Big\{ H_{SB}(t-s)\rho_{S}(t)\otimes\rho_{B}H_{SB}(t)-H_{SB}(t)H_{SB}(t-s)\rho_{S}(t)\otimes\rho_{B}\Big\}+\mathbf{h.c.}\label{eq:App-Mark}
\end{align}
Expanding the two terms in the above equation, we obtain (denoting
$\Delta_{mn}:=\varepsilon_{m}-\varepsilon_{n}$) 
\begin{align}
 & \mathrm{Tr}_{B}\int_{0}^{\infty}ds\,H_{SB}(t-s)\rho_{S}\otimes\rho_{B}H_{SB}(t)\label{eq:ME-term1}\\
= & \sum_{m,n=1}^{2}\int_{0}^{\infty}ds\,\hat{\tau}_{m}^{+}\rho_{S}\hat{\tau}_{n}^{-}\cdot e^{i\varepsilon_{m}(t-s)}e^{-i\varepsilon_{n}t}\cdot\langle\hat{B}_{n}^{\dagger}(t)\hat{B}_{m}(t-s)\rangle+\hat{\tau}_{m}^{-}\rho_{S}\hat{\tau}_{n}^{+}\cdot e^{-i\varepsilon_{m}(t-s)}e^{i\varepsilon_{n}t}\cdot\langle\hat{B}_{n}(t)\hat{B}_{m}^{\dagger}(t-s)\rangle\nonumber \\
= & \sum_{mn}\int_{0}^{\infty}ds\int_{0}^{\infty}\frac{d\nu}{2\pi}\,\hat{\tau}_{m}^{+}\rho_{S}\hat{\tau}_{n}^{-}e^{i\Delta_{mn}t}\cdot e^{-i(\varepsilon_{m}-\nu)s}\cdot J_{nm}(\nu)\tilde{n}(\nu)+\hat{\tau}_{m}^{-}\rho_{S}\hat{\tau}_{n}^{+}e^{-i\Delta_{mn}t}\cdot e^{i(\varepsilon_{m}-\nu)s}\cdot J_{mn}(\nu)[\tilde{n}(\nu)+1]\nonumber \\
\simeq & \sum_{mn}\frac{J_{nm}(\varepsilon_{m})}{2}\tilde{n}(\varepsilon_{m})\cdot\hat{\tau}_{m}^{+}\rho_{S}\hat{\tau}_{n}^{-}e^{i\Delta_{mn}t}+\frac{J_{mn}(\varepsilon_{m})}{2}[\tilde{n}(\varepsilon_{m})+1]\cdot\hat{\tau}_{m}^{-}\rho_{S}\hat{\tau}_{n}^{+}e^{-i\Delta_{mn}t}\nonumber 
\end{align}
\begin{align}
 & \mathrm{Tr}_{B}\int_{0}^{\infty}ds\,H_{SB}(t)H_{SB}(t-s)\rho_{S}(t)\otimes\rho_{B}\label{eq:ME-term2}\\
= & \sum_{m,n=1}^{2}\int_{0}^{\infty}ds\,\hat{\tau}_{m}^{+}\hat{\tau}_{n}^{-}\rho_{S}\cdot e^{i\varepsilon_{m}t}e^{-i\varepsilon_{n}(t-s)}\cdot\langle\hat{B}_{m}(t)\hat{B}_{n}^{\dagger}(t-s)\rangle+\hat{\tau}_{m}^{-}\hat{\tau}_{n}^{+}\rho_{S}\cdot e^{-i\varepsilon_{m}t}e^{i\varepsilon_{n}(t-s)}\cdot\langle\hat{B}_{m}^{\dagger}(t)\hat{B}_{n}(t-s)\rangle\nonumber \\
= & \sum_{mn}\int_{0}^{\infty}ds\int_{0}^{\infty}\frac{d\nu}{2\pi}\,\hat{\tau}_{m}^{+}\hat{\tau}_{n}^{-}\rho_{S}\cdot e^{i\Delta_{mn}t}e^{i(\varepsilon_{n}-\nu)s}\cdot J_{nm}(\nu)[\tilde{n}(\nu)+1]+\hat{\tau}_{m}^{-}\hat{\tau}_{n}^{+}\rho_{S}\cdot e^{-i\Delta_{mn}t}e^{-i(\varepsilon_{n}-\nu)s}\cdot J_{mn}(\nu)\tilde{n}(\nu)\nonumber \\
\simeq & \sum_{mn}\frac{J_{nm}(\varepsilon_{n})}{2}[\tilde{n}(\varepsilon_{n})+1]\cdot\hat{\tau}_{m}^{+}\hat{\tau}_{n}^{-}\rho_{S}e^{i\Delta_{mn}t}+\frac{J_{mn}(\varepsilon_{n})}{2}\tilde{n}(\varepsilon_{n})\cdot\hat{\tau}_{m}^{-}\hat{\tau}_{n}^{+}\rho_{S}e^{-i\Delta_{mn}t}\nonumber 
\end{align}
 In the above calculation, we define the coupling spectrum $J_{mn}(\omega)$
as
\begin{equation}
J_{mn}(\omega):=2\pi\sum_{k}g_{m,k}^{*}g_{n,k}\delta(\omega-\omega_{k})=[J_{nm}(\omega)]^{*},
\end{equation}
so the bath correlation functions are
\begin{align}
\langle\hat{B}_{n}^{\dagger}(t)\hat{B}_{m}(t-s)\rangle & =\sum_{k}g_{n,k}^{*}g_{m,k}\langle\hat{b}_{k}^{\dagger}\hat{b}_{k}\rangle e^{i\nu s}=\int_{0}^{\infty}\frac{d\nu}{2\pi}\,J_{nm}(\nu)\tilde{n}(\nu)e^{i\nu s},\nonumber \\
\langle\hat{B}_{n}(t)\hat{B}_{m}^{\dagger}(t-s)\rangle & =\sum_{k}g_{m,k}^{*}g_{n,k}\langle\hat{b}_{k}\hat{b}_{k}^{\dagger}\rangle e^{-i\nu s}=\int_{0}^{\infty}\frac{d\nu}{2\pi}\,J_{mn}(\nu)[\tilde{n}(\nu)+1]e^{-i\nu s}.
\end{align}
Notice that since the non-thermal bath state has the form of Eq.\,(\ref{eq:Bath-state}),
we have $\mathrm{Tr}[\rho_{B}\cdot\hat{b}_{k}^{\dagger}\hat{b}_{q}]=\delta_{kq}\cdot\overline{n}_{\mathrm{p}}(\omega_{k},T_{k})$,
and so we use $\tilde{n}(\omega_{k}):=\langle\hat{b}_{k}^{\dagger}\hat{b}_{k}\rangle$
to replace the standard Planck distribution $\overline{n}_{\mathrm{p}}(\omega,T)$
in thermal baths. For example, a temperature distribution like Eq.\,(\ref{eq:temp-dist})
would lead to 
\begin{equation}
\tilde{n}(\omega)=\begin{cases}
\cfrac{1}{\exp(\omega/T_{1})-1}, & 0<\omega\le\overline{\varepsilon}\\
\cfrac{1}{\exp(\omega/T_{2})-1}, & \omega>\overline{\varepsilon}
\end{cases}
\end{equation}
In the calculation of Eqs.\,(\ref{eq:ME-term1}, \ref{eq:ME-term2}),
we have utilized the formula
\begin{equation}
\int_{0}^{\infty}e^{i\omega s}ds=\pi\delta(\omega)+i\mathbf{P}\frac{1}{\omega},
\end{equation}
and the principal integrals in the the imaginary parts are omitted.

Therefore, in the Schr\"odinger's picture, we obtain
\begin{align}
\dot{\rho}_{S}=i[\rho_{S},H_{S}]+ & \sum_{m,n=1}^{2}\frac{1}{2}J_{nm}(\omega_{m})\tilde{n}(\omega_{m})\cdot[\hat{\tau}_{m}^{+}\rho_{S},\,\hat{\tau}_{n}^{-}]+\mathbf{h.c.}\nonumber \\
 & +\frac{1}{2}J_{mn}(\omega_{m})[\tilde{n}(\omega_{m})+1]\cdot[\hat{\tau}_{m}^{-}\rho_{S},\,\hat{\tau}_{n}^{+}]+\mathbf{h.c.}\label{eq:App-Mme}
\end{align}
Defining $\Gamma_{mn}^{+}(\omega):=J_{nm}(\omega)\tilde{n}(\omega)$,
$\Gamma_{mn}^{-}(\omega):=J_{mn}(\omega)[\tilde{n}(\omega)+1]$, and
$D_{mn}^{\pm}:=[\Gamma_{mn}^{\pm}(\varepsilon_{m})+\Gamma_{mn}^{\pm}(\varepsilon_{n})]/2$,
we can verify that the above master equation can be written in the
standard Lindblad form as Eq.\,(\ref{eq:ME-mark}) with a unitary
correction term Eq.\,(\ref{eq:H_c}).

To derive the time-dependent master equation (\ref{eq:ME-non-Mark}),
we can change the integral upper limit in Eqs.\,(\ref{eq:App-Mark},
\ref{eq:ME-term1}, \ref{eq:ME-term2}). Thus, we obtain 
\begin{align}
 & \mathrm{Tr}_{B}\int_{0}^{t}ds\,H_{SB}(t-s)\rho_{S}\otimes\rho_{B}H_{SB}(t)\\
= & \sum_{mn}\Big[\int_{0}^{t}ds\int_{0}^{\infty}\frac{d\nu}{2\pi}\,e^{-i(\varepsilon_{m}-\nu)s}J_{nm}(\nu)\tilde{n}(\nu)\Big]\hat{\tau}_{m}^{+}\rho_{S}\hat{\tau}_{n}^{-}e^{i\Delta_{mn}t}+\Big[\int_{0}^{t}ds\int_{0}^{\infty}\frac{d\nu}{2\pi}\,e^{i(\varepsilon_{m}-\nu)s}J_{mn}(\nu)[\tilde{n}(\nu)+1]\Big]\hat{\tau}_{m}^{-}\rho_{S}\hat{\tau}_{n}^{+}e^{-i\Delta_{mn}t},\nonumber 
\end{align}
\begin{align}
 & \mathrm{Tr}_{B}\int_{0}^{t}ds\,H_{SB}(t)H_{SB}(t-s)\rho_{S}(t)\otimes\rho_{B}\\
= & \sum_{mn}\Big[\int_{0}^{t}ds\int_{0}^{\infty}\frac{d\nu}{2\pi}\,e^{i(\varepsilon_{n}-\nu)s}J_{nm}(\nu)[\tilde{n}(\nu)+1]\Big]\hat{\tau}_{m}^{+}\hat{\tau}_{n}^{-}\rho_{S}e^{i\Delta_{mn}t}+\Big[\int_{0}^{t}ds\int_{0}^{\infty}\frac{d\nu}{2\pi}\,e^{-i(\varepsilon_{n}-\nu)s}J_{mn}(\nu)\tilde{n}(\nu)\Big]\hat{\tau}_{m}^{-}\hat{\tau}_{n}^{+}\rho_{S}e^{-i\Delta_{mn}t}.\nonumber 
\end{align}
 The coefficients become time-dependent, so we define the decay rates
as {[}Eq.\,(\ref{eq:Gam(w,t)}){]}
\begin{align}
\Gamma_{mn}^{+}(\omega,t): & =\Re\mathrm{e}\int_{0}^{t}ds\int_{0}^{\infty}\frac{d\nu}{2\pi}\,J_{nm}(\nu)\tilde{n}(\nu)e^{i(\nu-\omega)s},\\
\Gamma_{mn}^{-}(\omega,t): & =\Re\mathrm{e}\int_{0}^{t}ds\int_{0}^{\infty}\frac{d\nu}{2\pi}\,J_{mn}(\nu)[\tilde{n}(\nu)+1]e^{-i(\nu-\omega)s}.\nonumber 
\end{align}
Here we also omit the imaginary part, so $\Gamma_{mn}^{\pm}(\omega,t)$
will return to the homogenous Markovian case directly when $t\rightarrow\infty$.
With these changes, the time-dependent master equation is
\begin{align}
\dot{\rho}_{S}=i[\rho_{S},H_{S}]+ & \sum_{m,n=1}^{2}\frac{1}{2}\Gamma_{mn}^{+}(\varepsilon_{m},t)\cdot[\hat{\tau}_{m}^{+}\rho_{S},\,\hat{\tau}_{n}^{-}]+\mathbf{h.c.}\nonumber \\
 & +\frac{1}{2}\Gamma_{mn}^{-}(\varepsilon_{m},t)\cdot[\hat{\tau}_{m}^{-}\rho_{S},\,\hat{\tau}_{n}^{+}]+\mathbf{h.c.}
\end{align}
which is similar to Eq.\,(\ref{eq:App-Mme}). We can verify that
this master equation can be also written in the form of Eq.\,(\ref{eq:ME-non-Mark}).

\end{widetext}

\bibliographystyle{apsrev4-1}
\bibliography{Refs}

\begin{thebibliography}{34}%
\makeatletter
\providecommand \@ifxundefined [1]{%
 \@ifx{#1\undefined}
}%
\providecommand \@ifnum [1]{%
 \ifnum #1\expandafter \@firstoftwo
 \else \expandafter \@secondoftwo
 \fi
}%
\providecommand \@ifx [1]{%
 \ifx #1\expandafter \@firstoftwo
 \else \expandafter \@secondoftwo
 \fi
}%
\providecommand \natexlab [1]{#1}%
\providecommand \enquote  [1]{``#1''}%
\providecommand \bibnamefont  [1]{#1}%
\providecommand \bibfnamefont [1]{#1}%
\providecommand \citenamefont [1]{#1}%
\providecommand \href@noop [0]{\@secondoftwo}%
\providecommand \href [0]{\begingroup \@sanitize@url \@href}%
\providecommand \@href[1]{\@@startlink{#1}\@@href}%
\providecommand \@@href[1]{\endgroup#1\@@endlink}%
\providecommand \@sanitize@url [0]{\catcode `\\12\catcode `\$12\catcode
  `\&12\catcode `\#12\catcode `\^12\catcode `\_12\catcode `\%12\relax}%
\providecommand \@@startlink[1]{}%
\providecommand \@@endlink[0]{}%
\providecommand \url  [0]{\begingroup\@sanitize@url \@url }%
\providecommand \@url [1]{\endgroup\@href {#1}{\urlprefix }}%
\providecommand \urlprefix  [0]{URL }%
\providecommand \Eprint [0]{\href }%
\providecommand \doibase [0]{http://dx.doi.org/}%
\providecommand \selectlanguage [0]{\@gobble}%
\providecommand \bibinfo  [0]{\@secondoftwo}%
\providecommand \bibfield  [0]{\@secondoftwo}%
\providecommand \translation [1]{[#1]}%
\providecommand \BibitemOpen [0]{}%
\providecommand \bibitemStop [0]{}%
\providecommand \bibitemNoStop [0]{.\EOS\space}%
\providecommand \EOS [0]{\spacefactor3000\relax}%
\providecommand \BibitemShut  [1]{\csname bibitem#1\endcsname}%
\let\auto@bib@innerbib\@empty
\bibitem [{\citenamefont {Duan}\ and\ \citenamefont
  {Guo}(1997)}]{duan_preserving_1997}%
  \BibitemOpen
  \bibfield  {author} {\bibinfo {author} {\bibfnamefont {L.-M.}\ \bibnamefont
  {Duan}}\ and\ \bibinfo {author} {\bibfnamefont {G.-C.}\ \bibnamefont {Guo}},\
  }\href {\doibase 10.1103/PhysRevLett.79.1953} {\bibfield  {journal} {\bibinfo
   {journal} {Phys. Rev. Lett.}\ }\textbf {\bibinfo {volume} {79}},\ \bibinfo
  {pages} {1953} (\bibinfo {year} {1997})}\BibitemShut {NoStop}%
\bibitem [{\citenamefont {Zanardi}\ and\ \citenamefont
  {Rasetti}(1997)}]{zanardi_noiseless_1997}%
  \BibitemOpen
  \bibfield  {author} {\bibinfo {author} {\bibfnamefont {P.}~\bibnamefont
  {Zanardi}}\ and\ \bibinfo {author} {\bibfnamefont {M.}~\bibnamefont
  {Rasetti}},\ }\href {\doibase 10.1103/PhysRevLett.79.3306} {\bibfield
  {journal} {\bibinfo  {journal} {Phys. Rev. Lett.}\ }\textbf {\bibinfo
  {volume} {79}},\ \bibinfo {pages} {3306} (\bibinfo {year}
  {1997})}\BibitemShut {NoStop}%
\bibitem [{\citenamefont {Lidar}\ \emph {et~al.}(1998)\citenamefont {Lidar},
  \citenamefont {Chuang},\ and\ \citenamefont
  {Whaley}}]{lidar_decoherence-free_1998}%
  \BibitemOpen
  \bibfield  {author} {\bibinfo {author} {\bibfnamefont {D.~A.}\ \bibnamefont
  {Lidar}}, \bibinfo {author} {\bibfnamefont {I.~L.}\ \bibnamefont {Chuang}}, \
  and\ \bibinfo {author} {\bibfnamefont {K.~B.}\ \bibnamefont {Whaley}},\
  }\href {\doibase 10.1103/PhysRevLett.81.2594} {\bibfield  {journal} {\bibinfo
   {journal} {Phys. Rev. Lett.}\ }\textbf {\bibinfo {volume} {81}},\ \bibinfo
  {pages} {2594} (\bibinfo {year} {1998})}\BibitemShut {NoStop}%
\bibitem [{\citenamefont {Scully}\ and\ \citenamefont
  {Zubairy}(1997)}]{scully_quantum_1997}%
  \BibitemOpen
  \bibfield  {author} {\bibinfo {author} {\bibfnamefont {M.~O.}\ \bibnamefont
  {Scully}}\ and\ \bibinfo {author} {\bibfnamefont {M.~S.}\ \bibnamefont
  {Zubairy}},\ }\href@noop {} {\emph {\bibinfo {title} {Quantum optics}}}\
  (\bibinfo  {publisher} {Cambridge university press},\ \bibinfo {year}
  {1997})\BibitemShut {NoStop}%
\bibitem [{\citenamefont {Ro{\ss}nagel}\ \emph {et~al.}(2014)\citenamefont
  {Ro{\ss}nagel}, \citenamefont {Abah}, \citenamefont {Schmidt-Kaler},
  \citenamefont {Singer},\ and\ \citenamefont
  {Lutz}}]{rosnagel_nanoscale_2014}%
  \BibitemOpen
  \bibfield  {author} {\bibinfo {author} {\bibfnamefont {J.}~\bibnamefont
  {Ro{\ss}nagel}}, \bibinfo {author} {\bibfnamefont {O.}~\bibnamefont {Abah}},
  \bibinfo {author} {\bibfnamefont {F.}~\bibnamefont {Schmidt-Kaler}}, \bibinfo
  {author} {\bibfnamefont {K.}~\bibnamefont {Singer}}, \ and\ \bibinfo {author}
  {\bibfnamefont {E.}~\bibnamefont {Lutz}},\ }\href {\doibase
  10.1103/PhysRevLett.112.030602} {\bibfield  {journal} {\bibinfo  {journal}
  {Phys. Rev. Lett.}\ }\textbf {\bibinfo {volume} {112}},\ \bibinfo {pages}
  {030602} (\bibinfo {year} {2014})}\BibitemShut {NoStop}%
\bibitem [{\citenamefont {Alicki}\ and\ \citenamefont
  {Gelbwaser-Klimovsky}(2015)}]{alicki_non-equilibrium_2015}%
  \BibitemOpen
  \bibfield  {author} {\bibinfo {author} {\bibfnamefont {R.}~\bibnamefont
  {Alicki}}\ and\ \bibinfo {author} {\bibfnamefont {D.}~\bibnamefont
  {Gelbwaser-Klimovsky}},\ }\href {\doibase 10.1088/1367-2630/17/11/115012}
  {\bibfield  {journal} {\bibinfo  {journal} {New J. Phys.}\ }\textbf {\bibinfo
  {volume} {17}},\ \bibinfo {pages} {115012} (\bibinfo {year}
  {2015})}\BibitemShut {NoStop}%
\bibitem [{\citenamefont {Dorfman}\ \emph {et~al.}(2013)\citenamefont
  {Dorfman}, \citenamefont {Voronine}, \citenamefont {Mukamel},\ and\
  \citenamefont {Scully}}]{dorfman_photosynthetic_2013}%
  \BibitemOpen
  \bibfield  {author} {\bibinfo {author} {\bibfnamefont {K.~E.}\ \bibnamefont
  {Dorfman}}, \bibinfo {author} {\bibfnamefont {D.~V.}\ \bibnamefont
  {Voronine}}, \bibinfo {author} {\bibfnamefont {S.}~\bibnamefont {Mukamel}}, \
  and\ \bibinfo {author} {\bibfnamefont {M.~O.}\ \bibnamefont {Scully}},\
  }\href {\doibase 10.1073/pnas.1212666110} {\bibfield  {journal} {\bibinfo
  {journal} {Proc. Nat. Acad. Sci.}\ }\textbf {\bibinfo {volume} {110}},\
  \bibinfo {pages} {2746} (\bibinfo {year} {2013})}\BibitemShut {NoStop}%
\bibitem [{\citenamefont {Creatore}\ \emph {et~al.}(2013)\citenamefont
  {Creatore}, \citenamefont {Parker}, \citenamefont {Emmott},\ and\
  \citenamefont {Chin}}]{creatore_efficient_2013}%
  \BibitemOpen
  \bibfield  {author} {\bibinfo {author} {\bibfnamefont {C.}~\bibnamefont
  {Creatore}}, \bibinfo {author} {\bibfnamefont {M.~A.}\ \bibnamefont
  {Parker}}, \bibinfo {author} {\bibfnamefont {S.}~\bibnamefont {Emmott}}, \
  and\ \bibinfo {author} {\bibfnamefont {A.~W.}\ \bibnamefont {Chin}},\ }\href
  {\doibase 10.1103/PhysRevLett.111.253601} {\bibfield  {journal} {\bibinfo
  {journal} {Phys. Rev. Lett.}\ }\textbf {\bibinfo {volume} {111}},\ \bibinfo
  {pages} {253601} (\bibinfo {year} {2013})}\BibitemShut {NoStop}%
\bibitem [{\citenamefont {Ol{\v s}ina}\ \emph {et~al.}(2014)\citenamefont
  {Ol{\v s}ina}, \citenamefont {Dijkstra}, \citenamefont {Wang},\ and\
  \citenamefont {Cao}}]{olsina_can_2014}%
  \BibitemOpen
  \bibfield  {author} {\bibinfo {author} {\bibfnamefont {J.}~\bibnamefont
  {Ol{\v s}ina}}, \bibinfo {author} {\bibfnamefont {A.~G.}\ \bibnamefont
  {Dijkstra}}, \bibinfo {author} {\bibfnamefont {C.}~\bibnamefont {Wang}}, \
  and\ \bibinfo {author} {\bibfnamefont {J.}~\bibnamefont {Cao}},\ }\href
  {http://arxiv.org/abs/1408.5385} {\bibfield  {journal} {\bibinfo  {journal}
  {arXiv:1408.5385}\ } (\bibinfo {year} {2014})}\BibitemShut {NoStop}%
\bibitem [{\citenamefont {Hu}\ \emph {et~al.}(1992)\citenamefont {Hu},
  \citenamefont {Paz},\ and\ \citenamefont {Zhang}}]{hu_quantum_1992}%
  \BibitemOpen
  \bibfield  {author} {\bibinfo {author} {\bibfnamefont {B.~L.}\ \bibnamefont
  {Hu}}, \bibinfo {author} {\bibfnamefont {J.~P.}\ \bibnamefont {Paz}}, \ and\
  \bibinfo {author} {\bibfnamefont {Y.}~\bibnamefont {Zhang}},\ }\href
  {\doibase 10.1103/PhysRevD.45.2843} {\bibfield  {journal} {\bibinfo
  {journal} {Phys. Rev. D}\ }\textbf {\bibinfo {volume} {45}},\ \bibinfo
  {pages} {2843} (\bibinfo {year} {1992})}\BibitemShut {NoStop}%
\bibitem [{\citenamefont {Breuer}\ and\ \citenamefont
  {Petruccione}(2002)}]{breuer_theory_2002}%
  \BibitemOpen
  \bibfield  {author} {\bibinfo {author} {\bibfnamefont {H.}~\bibnamefont
  {Breuer}}\ and\ \bibinfo {author} {\bibfnamefont {F.}~\bibnamefont
  {Petruccione}},\ }\href@noop {} {\emph {\bibinfo {title} {The theory of open
  quantum systems}}}\ (\bibinfo  {publisher} {Oxford University Press},\
  \bibinfo {year} {2002})\BibitemShut {NoStop}%
\bibitem [{\citenamefont {Ma}\ \emph {et~al.}(2012)\citenamefont {Ma},
  \citenamefont {Sun}, \citenamefont {Wang},\ and\ \citenamefont
  {Nori}}]{ma_entanglement_2012}%
  \BibitemOpen
  \bibfield  {author} {\bibinfo {author} {\bibfnamefont {J.}~\bibnamefont
  {Ma}}, \bibinfo {author} {\bibfnamefont {Z.}~\bibnamefont {Sun}}, \bibinfo
  {author} {\bibfnamefont {X.}~\bibnamefont {Wang}}, \ and\ \bibinfo {author}
  {\bibfnamefont {F.}~\bibnamefont {Nori}},\ }\href {\doibase
  10.1103/PhysRevA.85.062323} {\bibfield  {journal} {\bibinfo  {journal} {Phys.
  Rev. A}\ }\textbf {\bibinfo {volume} {85}},\ \bibinfo {pages} {062323}
  (\bibinfo {year} {2012})}\BibitemShut {NoStop}%
\bibitem [{\citenamefont {Yang}\ \emph {et~al.}(2013)\citenamefont {Yang},
  \citenamefont {Cai}, \citenamefont {Xu}, \citenamefont {Zhang},\ and\
  \citenamefont {Sun}}]{yang_master_2013}%
  \BibitemOpen
  \bibfield  {author} {\bibinfo {author} {\bibfnamefont {L.-P.}\ \bibnamefont
  {Yang}}, \bibinfo {author} {\bibfnamefont {C.~Y.}\ \bibnamefont {Cai}},
  \bibinfo {author} {\bibfnamefont {D.~Z.}\ \bibnamefont {Xu}}, \bibinfo
  {author} {\bibfnamefont {W.-M.}\ \bibnamefont {Zhang}}, \ and\ \bibinfo
  {author} {\bibfnamefont {C.~P.}\ \bibnamefont {Sun}},\ }\href {\doibase
  10.1103/PhysRevA.87.012110} {\bibfield  {journal} {\bibinfo  {journal} {Phys.
  Rev. A}\ }\textbf {\bibinfo {volume} {87}},\ \bibinfo {pages} {012110}
  (\bibinfo {year} {2013})}\BibitemShut {NoStop}%
\bibitem [{\citenamefont {Li}\ \emph {et~al.}(2014{\natexlab{a}})\citenamefont
  {Li}, \citenamefont {Yip}, \citenamefont {Deng}, \citenamefont {Chen},
  \citenamefont {Yu}, \citenamefont {You},\ and\ \citenamefont
  {Lam}}]{li_approach_2014}%
  \BibitemOpen
  \bibfield  {author} {\bibinfo {author} {\bibfnamefont {Z.-Z.}\ \bibnamefont
  {Li}}, \bibinfo {author} {\bibfnamefont {C.-T.}\ \bibnamefont {Yip}},
  \bibinfo {author} {\bibfnamefont {H.-Y.}\ \bibnamefont {Deng}}, \bibinfo
  {author} {\bibfnamefont {M.}~\bibnamefont {Chen}}, \bibinfo {author}
  {\bibfnamefont {T.}~\bibnamefont {Yu}}, \bibinfo {author} {\bibfnamefont
  {J.~Q.}\ \bibnamefont {You}}, \ and\ \bibinfo {author} {\bibfnamefont
  {C.-H.}\ \bibnamefont {Lam}},\ }\href {\doibase 10.1103/PhysRevA.90.022122}
  {\bibfield  {journal} {\bibinfo  {journal} {Phys. Rev. A}\ }\textbf {\bibinfo
  {volume} {90}},\ \bibinfo {pages} {022122} (\bibinfo {year}
  {2014}{\natexlab{a}})}\BibitemShut {NoStop}%
\bibitem [{\citenamefont {Weiss}(2012)}]{weiss_quantum_2012}%
  \BibitemOpen
  \bibfield  {author} {\bibinfo {author} {\bibfnamefont {U.}~\bibnamefont
  {Weiss}},\ }\href@noop {} {\emph {\bibinfo {title} {Quantum dissipative
  systems}}}\ (\bibinfo  {publisher} {World Scientific},\ \bibinfo {year}
  {2012})\BibitemShut {NoStop}%
\bibitem [{\citenamefont {Agarwal}(1974)}]{agarwal_quantum_1974}%
  \BibitemOpen
  \bibfield  {author} {\bibinfo {author} {\bibfnamefont {G.~S.}\ \bibnamefont
  {Agarwal}},\ }\href@noop {} {\emph {\bibinfo {title} {Quantum statistical
  theories of spontaneous emission and their relation to other approaches}}}\
  (\bibinfo  {publisher} {Springer},\ \bibinfo {year} {1974})\BibitemShut
  {NoStop}%
\bibitem [{\citenamefont {Zhu}\ \emph {et~al.}(1995)\citenamefont {Zhu},
  \citenamefont {Narducci},\ and\ \citenamefont
  {Scully}}]{zhu_quantum-mechanical_1995}%
  \BibitemOpen
  \bibfield  {author} {\bibinfo {author} {\bibfnamefont {S.-Y.}\ \bibnamefont
  {Zhu}}, \bibinfo {author} {\bibfnamefont {L.~M.}\ \bibnamefont {Narducci}}, \
  and\ \bibinfo {author} {\bibfnamefont {M.~O.}\ \bibnamefont {Scully}},\
  }\href {\doibase 10.1103/PhysRevA.52.4791} {\bibfield  {journal} {\bibinfo
  {journal} {Phys. Rev. A}\ }\textbf {\bibinfo {volume} {52}},\ \bibinfo
  {pages} {4791} (\bibinfo {year} {1995})}\BibitemShut {NoStop}%
\bibitem [{\citenamefont {Li}\ \emph {et~al.}(2010)\citenamefont {Li},
  \citenamefont {Wang}, \citenamefont {Zheng}, \citenamefont {Zhu},\ and\
  \citenamefont {Zubairy}}]{li_quantum_2010}%
  \BibitemOpen
  \bibfield  {author} {\bibinfo {author} {\bibfnamefont {Z.-H.}\ \bibnamefont
  {Li}}, \bibinfo {author} {\bibfnamefont {D.-W.}\ \bibnamefont {Wang}},
  \bibinfo {author} {\bibfnamefont {H.}~\bibnamefont {Zheng}}, \bibinfo
  {author} {\bibfnamefont {S.-Y.}\ \bibnamefont {Zhu}}, \ and\ \bibinfo
  {author} {\bibfnamefont {M.~S.}\ \bibnamefont {Zubairy}},\ }\href {\doibase
  10.1103/PhysRevA.82.050501} {\bibfield  {journal} {\bibinfo  {journal} {Phys.
  Rev. A}\ }\textbf {\bibinfo {volume} {82}},\ \bibinfo {pages} {050501}
  (\bibinfo {year} {2010})}\BibitemShut {NoStop}%
\bibitem [{\citenamefont {Xu}\ \emph {et~al.}(2016)\citenamefont {Xu},
  \citenamefont {Wang}, \citenamefont {Zhao},\ and\ \citenamefont
  {Cao}}]{xu_polaron_2016}%
  \BibitemOpen
  \bibfield  {author} {\bibinfo {author} {\bibfnamefont {D.}~\bibnamefont
  {Xu}}, \bibinfo {author} {\bibfnamefont {C.}~\bibnamefont {Wang}}, \bibinfo
  {author} {\bibfnamefont {Y.}~\bibnamefont {Zhao}}, \ and\ \bibinfo {author}
  {\bibfnamefont {J.}~\bibnamefont {Cao}},\ }\href {\doibase
  10.1088/1367-2630/18/2/023003} {\bibfield  {journal} {\bibinfo  {journal}
  {New J. Phys.}\ }\textbf {\bibinfo {volume} {18}},\ \bibinfo {pages} {023003}
  (\bibinfo {year} {2016})}\BibitemShut {NoStop}%
\bibitem [{\citenamefont {Orszag}(2000)}]{orszag_quantum_2000}%
  \BibitemOpen
  \bibfield  {author} {\bibinfo {author} {\bibfnamefont {M.}~\bibnamefont
  {Orszag}},\ }\href@noop {} {\emph {\bibinfo {title} {Quantum optics}}}\
  (\bibinfo  {publisher} {Springer},\ \bibinfo {year} {2000})\BibitemShut
  {NoStop}%
\bibitem [{\citenamefont {Li}\ \emph {et~al.}(2014{\natexlab{b}})\citenamefont
  {Li}, \citenamefont {Yang},\ and\ \citenamefont {Sun}}]{li_long-term_2014}%
  \BibitemOpen
  \bibfield  {author} {\bibinfo {author} {\bibfnamefont {S.-W.}\ \bibnamefont
  {Li}}, \bibinfo {author} {\bibfnamefont {L.-P.}\ \bibnamefont {Yang}}, \ and\
  \bibinfo {author} {\bibfnamefont {C.-P.}\ \bibnamefont {Sun}},\ }\href
  {\doibase 10.1140/epjd/e2014-40659-8} {\bibfield  {journal} {\bibinfo
  {journal} {Eur. Phys. J. D}\ }\textbf {\bibinfo {volume} {68}},\ \bibinfo
  {pages} {45} (\bibinfo {year} {2014}{\natexlab{b}})},\ \bibinfo {note}
  {arXiv:1303.1266}\BibitemShut {NoStop}%
\bibitem [{\citenamefont {Li}\ \emph {et~al.}(2015)\citenamefont {Li},
  \citenamefont {Cai},\ and\ \citenamefont {Sun}}]{li_steady_2015}%
  \BibitemOpen
  \bibfield  {author} {\bibinfo {author} {\bibfnamefont {S.-W.}\ \bibnamefont
  {Li}}, \bibinfo {author} {\bibfnamefont {C.~Y.}\ \bibnamefont {Cai}}, \ and\
  \bibinfo {author} {\bibfnamefont {C.~P.}\ \bibnamefont {Sun}},\ }\href
  {\doibase 10.1016/j.aop.2015.05.004} {\bibfield  {journal} {\bibinfo
  {journal} {Ann. Phys.}\ }\textbf {\bibinfo {volume} {360}},\ \bibinfo {pages}
  {19} (\bibinfo {year} {2015})}\BibitemShut {NoStop}%
\bibitem [{\citenamefont {Wei}(1994)}]{wei_time_1994}%
  \BibitemOpen
  \bibfield  {author} {\bibinfo {author} {\bibfnamefont {W.~W.-S.}\
  \bibnamefont {Wei}},\ }\href@noop {} {\emph {\bibinfo {title} {Time series
  analysis}}}\ (\bibinfo  {publisher} {Addison-Wesley Redwood City,
  California},\ \bibinfo {year} {1994})\BibitemShut {NoStop}%
\bibitem [{\citenamefont {Wichterich}\ \emph {et~al.}(2007)\citenamefont
  {Wichterich}, \citenamefont {Henrich}, \citenamefont {Breuer}, \citenamefont
  {Gemmer},\ and\ \citenamefont {Michel}}]{wichterich_modeling_2007}%
  \BibitemOpen
  \bibfield  {author} {\bibinfo {author} {\bibfnamefont {H.}~\bibnamefont
  {Wichterich}}, \bibinfo {author} {\bibfnamefont {M.~J.}\ \bibnamefont
  {Henrich}}, \bibinfo {author} {\bibfnamefont {H.-P.}\ \bibnamefont {Breuer}},
  \bibinfo {author} {\bibfnamefont {J.}~\bibnamefont {Gemmer}}, \ and\ \bibinfo
  {author} {\bibfnamefont {M.}~\bibnamefont {Michel}},\ }\href {\doibase
  10.1103/PhysRevE.76.031115} {\bibfield  {journal} {\bibinfo  {journal} {Phys.
  Rev. E}\ }\textbf {\bibinfo {volume} {76}},\ \bibinfo {pages} {031115}
  (\bibinfo {year} {2007})}\BibitemShut {NoStop}%
\bibitem [{\citenamefont {Bergmann}\ and\ \citenamefont
  {Lebowitz}(1955)}]{bergmann_new_1955}%
  \BibitemOpen
  \bibfield  {author} {\bibinfo {author} {\bibfnamefont {P.~G.}\ \bibnamefont
  {Bergmann}}\ and\ \bibinfo {author} {\bibfnamefont {J.~L.}\ \bibnamefont
  {Lebowitz}},\ }\href {\doibase 10.1103/PhysRev.99.578} {\bibfield  {journal}
  {\bibinfo  {journal} {Phys. Rev.}\ }\textbf {\bibinfo {volume} {99}},\
  \bibinfo {pages} {578} (\bibinfo {year} {1955})}\BibitemShut {NoStop}%
\bibitem [{\citenamefont {Cai}\ \emph {et~al.}(2014)\citenamefont {Cai},
  \citenamefont {Li}, \citenamefont {Liu},\ and\ \citenamefont
  {Sun}}]{cai_entropy_2014}%
  \BibitemOpen
  \bibfield  {author} {\bibinfo {author} {\bibfnamefont {C.-Y.}\ \bibnamefont
  {Cai}}, \bibinfo {author} {\bibfnamefont {S.-W.}\ \bibnamefont {Li}},
  \bibinfo {author} {\bibfnamefont {X.-F.}\ \bibnamefont {Liu}}, \ and\
  \bibinfo {author} {\bibfnamefont {C.~P.}\ \bibnamefont {Sun}},\ }\href
  {http://arxiv.org/abs/1407.2004} {\bibfield  {journal} {\bibinfo  {journal}
  {arXiv:1407.2004}\ } (\bibinfo {year} {2014})}\BibitemShut {NoStop}%
\bibitem [{\citenamefont {Su{\'a}rez}\ \emph {et~al.}(1992)\citenamefont
  {Su{\'a}rez}, \citenamefont {Silbey},\ and\ \citenamefont
  {Oppenheim}}]{suarez_memory_1992}%
  \BibitemOpen
  \bibfield  {author} {\bibinfo {author} {\bibfnamefont {A.}~\bibnamefont
  {Su{\'a}rez}}, \bibinfo {author} {\bibfnamefont {R.}~\bibnamefont {Silbey}},
  \ and\ \bibinfo {author} {\bibfnamefont {I.}~\bibnamefont {Oppenheim}},\
  }\href {\doibase 10.1063/1.463831} {\bibfield  {journal} {\bibinfo  {journal}
  {J. Chem. Phys.}\ }\textbf {\bibinfo {volume} {97}},\ \bibinfo {pages} {5101}
  (\bibinfo {year} {1992})}\BibitemShut {NoStop}%
\bibitem [{\citenamefont {Gorini}\ \emph {et~al.}(1976)\citenamefont {Gorini},
  \citenamefont {Kossakowski},\ and\ \citenamefont
  {Sudarshan}}]{gorini_completely_1976}%
  \BibitemOpen
  \bibfield  {author} {\bibinfo {author} {\bibfnamefont {V.}~\bibnamefont
  {Gorini}}, \bibinfo {author} {\bibfnamefont {A.}~\bibnamefont {Kossakowski}},
  \ and\ \bibinfo {author} {\bibfnamefont {E.~C.~G.}\ \bibnamefont
  {Sudarshan}},\ }\href {\doibase 10.1063/1.522979} {\bibfield  {journal}
  {\bibinfo  {journal} {J. Math. Phys.}\ }\textbf {\bibinfo {volume} {17}},\
  \bibinfo {pages} {821} (\bibinfo {year} {1976})}\BibitemShut {NoStop}%
\bibitem [{\citenamefont {Lindblad}(1976)}]{lindblad_generators_1976}%
  \BibitemOpen
  \bibfield  {author} {\bibinfo {author} {\bibfnamefont {G.}~\bibnamefont
  {Lindblad}},\ }\href {\doibase 10.1007/BF01608499} {\bibfield  {journal}
  {\bibinfo  {journal} {Comm. Math. Phys.}\ }\textbf {\bibinfo {volume} {48}},\
  \bibinfo {pages} {119} (\bibinfo {year} {1976})}\BibitemShut {NoStop}%
\bibitem [{\citenamefont {Clerk}\ \emph {et~al.}(2010)\citenamefont {Clerk},
  \citenamefont {Devoret}, \citenamefont {Girvin}, \citenamefont {Marquardt},\
  and\ \citenamefont {Schoelkopf}}]{clerk_introduction_2010}%
  \BibitemOpen
  \bibfield  {author} {\bibinfo {author} {\bibfnamefont {A.~A.}\ \bibnamefont
  {Clerk}}, \bibinfo {author} {\bibfnamefont {M.~H.}\ \bibnamefont {Devoret}},
  \bibinfo {author} {\bibfnamefont {S.~M.}\ \bibnamefont {Girvin}}, \bibinfo
  {author} {\bibfnamefont {F.}~\bibnamefont {Marquardt}}, \ and\ \bibinfo
  {author} {\bibfnamefont {R.~J.}\ \bibnamefont {Schoelkopf}},\ }\href
  {\doibase 10.1103/RevModPhys.82.1155} {\bibfield  {journal} {\bibinfo
  {journal} {Rev. Mod. Phys.}\ }\textbf {\bibinfo {volume} {82}},\ \bibinfo
  {pages} {1155} (\bibinfo {year} {2010})}\BibitemShut {NoStop}%
\bibitem [{\citenamefont {Gardiner}\ and\ \citenamefont
  {Zoller}(2004)}]{gardiner_quantum_2004}%
  \BibitemOpen
  \bibfield  {author} {\bibinfo {author} {\bibfnamefont {C.}~\bibnamefont
  {Gardiner}}\ and\ \bibinfo {author} {\bibfnamefont {P.}~\bibnamefont
  {Zoller}},\ }\href@noop {} {\emph {\bibinfo {title} {Quantum noise}}},\
  Vol.~\bibinfo {volume} {56}\ (\bibinfo  {publisher} {Springer},\ \bibinfo
  {year} {2004})\BibitemShut {NoStop}%
\bibitem [{\citenamefont {Rivas}\ \emph {et~al.}(2014)\citenamefont {Rivas},
  \citenamefont {Huelga},\ and\ \citenamefont {Plenio}}]{rivas_quantum_2014}%
  \BibitemOpen
  \bibfield  {author} {\bibinfo {author} {\bibfnamefont {{\'A}.}~\bibnamefont
  {Rivas}}, \bibinfo {author} {\bibfnamefont {S.~F.}\ \bibnamefont {Huelga}}, \
  and\ \bibinfo {author} {\bibfnamefont {M.~B.}\ \bibnamefont {Plenio}},\
  }\href {\doibase 10.1088/0034-4885/77/9/094001} {\bibfield  {journal}
  {\bibinfo  {journal} {Rep. Prog. Phys.}\ }\textbf {\bibinfo {volume} {77}},\
  \bibinfo {pages} {094001} (\bibinfo {year} {2014})}\BibitemShut {NoStop}%
\bibitem [{\citenamefont {Chru{\'s}ci{\'n}ski}\ and\ \citenamefont
  {Kossakowski}(2012)}]{chruscinski_markovianity_2012}%
  \BibitemOpen
  \bibfield  {author} {\bibinfo {author} {\bibfnamefont {D.}~\bibnamefont
  {Chru{\'s}ci{\'n}ski}}\ and\ \bibinfo {author} {\bibfnamefont
  {A.}~\bibnamefont {Kossakowski}},\ }\href {\doibase
  10.1088/0953-4075/45/15/154002} {\bibfield  {journal} {\bibinfo  {journal}
  {J. Phys. B}\ }\textbf {\bibinfo {volume} {45}},\ \bibinfo {pages} {154002}
  (\bibinfo {year} {2012})}\BibitemShut {NoStop}%
\bibitem [{\citenamefont {Hall}(2008)}]{hall_complete_2008}%
  \BibitemOpen
  \bibfield  {author} {\bibinfo {author} {\bibfnamefont {M.~J.~W.}\
  \bibnamefont {Hall}},\ }\href {\doibase 10.1088/1751-8113/41/20/205302}
  {\bibfield  {journal} {\bibinfo  {journal} {J. Phys. A}\ }\textbf {\bibinfo
  {volume} {41}},\ \bibinfo {pages} {205302} (\bibinfo {year}
  {2008})}\BibitemShut {NoStop}%
\end{thebibliography}%

\end{document}